\documentclass[doublespacing,reviewcopy,authoryear,12pt]{elsarticle}

\usepackage{amssymb}
\usepackage{amsmath} 
\usepackage{amsfonts} 
\usepackage{amssymb}  
\usepackage{color}
\usepackage{multirow}
\usepackage{graphicx}
\usepackage{topcapt}
\usepackage{multirow}
\usepackage{subfigure}
\usepackage{rotating}
\usepackage{supertabular}
\usepackage{booktabs}
\usepackage[large]{caption}
\usepackage{lineno}
\usepackage{setspace}
\journal{Icarus}

\begin{document}

\begin{frontmatter}


\title{The escape of heavy atoms from the ionosphere of HD209458b. II. Interpretation of the observations}

\author[rvt]{T. T. Koskinen\corref{cor1}}
\ead{tommi@lpl.arizona.edu}
\cortext[cor1]{Corresponding author. Fax: +1 (520) 621 4933}
\address[rvt]{Lunar and Planetary Laboratory, University of Arizona, 1629 E. University Blvd., Tucson, AZ 85721, USA}

\author[rvt]{R. V. Yelle}

\author[ucl]{M. J. Harris}
\address[ucl]{Department of Physics and Astronomy, University College London, Gower Street, London WC1E 6BT, UK}

\author[france]{P. Lavvas}
\address[france]{Groupe de Spectrom\'etrie Mol\'eculaire et Atmosph\'erique UMR CNRS 6089, Universit\'e Reims Champagne-Ardenne, 51687, France}

\begin{abstract}
Transits in the H I 1216~\AA~(Lyman $\alpha$), O I 1334~\AA, C II 1335~\AA, and Si III 1206.5~\AA~lines constrain the properties of the upper atmosphere of HD209458b.  In addition to probing the temperature and density profiles in the thermosphere, they have implications for the properties of the lower atmosphere.  Fits to the observations with a simple empirical model and a direct comparison with a more complex hydrodynamic model constrain the mean temperature and ionization state of the atmosphere, and imply that the optical depth of the extended thermosphere of the planet in the atomic resonance lines is significant.  In particular, it is sufficient to explain the observed transit depths in the H I 1216~\AA~line.  The detection of O at high altitudes implies that the minimum mass loss rate from the planet is approximately 6~$\times$~10$^6$ kg~s$^{-1}$.  The mass loss rate based on our hydrodynamic model is higher than this and implies that diffusive separation is prevented for neutral species with a mass lower than about 130 amu by the escape of H.  Heavy ions are transported to the upper atmosphere by Coulomb collisions with H$^+$ and their presence does not provide as strong constraints on the mass loss rate as the detection of heavy neutral atoms.  Models of the upper atmosphere with solar composition and heating based on the average solar X-ray and EUV flux agree broadly with the observations but tend to underestimate the transit depths in the O I, C II, and Si III lines.  This suggests that the temperature and/or elemental abundances in the thermosphere may be higher than expected from such models.  Observations of the escaping atmosphere can potentially be used to constrain the strength of the planetary magnetic field.  We find that a magnetic moment of $m \lesssim$~0.04 $m_J$, where $m_J$ is the Jovian magnetic moment, allows the ions to escape globally rather than only along open field lines.  The detection of Si$^{2+}$ in the thermosphere indicates that clouds of forsterite and enstatite do not form in the lower atmosphere.  This has implications for the temperature and dynamics of the atmosphere that also affect the interpretation of transit and secondary eclipse observations in the visible and infrared wavelengths.
\end{abstract}


\end{frontmatter}

\linenumbers

\section{Introduction}
\label{sc:intro}       

The detection of H, O, C$^+$, and Si$^{2+}$ in the upper atmosphere of HD209458b \citep{vidalmadjar03,vidalmadjar04,linsky10}, and the tentative detection of H in the upper atmosphere of HD189733b \citep{lecavelier10} and Mg$^+$ in the upper atmosphere of WASP-12b \citep{fossati10} are among the most exciting recent discoveries related to the atmospheres of extrasolar giant planets (EGPs).  The observations demonstrate that the the upper atmospheres of close-in EGPs such as HD209458b differ significantly from the thermospheres of the giant planets in the solar system.  They are much hotter, they extend to several planetary radii and instead of molecular hydrogen, they are primarily composed of atoms and atomic ions.  

The detection of heavy atoms and ions such as O, C$^+$, Si$^{2+}$, and Mg$^+$ implies that the atmospheres of close-in EGPs are not always affected by diffusive separation.  A likely explanation is that diffusive separation of the heavy atoms and ions is prevented by momentum transfer collisions with the rapidly escaping light atoms and ions.  Mass fractionation during hydrodynamic escape is believed to have played an important role in the early evolution of the atmospheres of the terrestrial planets \citep[e.g.,][]{zahnle86,hunten87} but it cannot be observed in action anywhere in the solar system.  Observations of EGP atmospheres thus provide a unique opportunity to study this phenomenon that should lead to a better understanding of evolutionary processes in our own solar system.    

Extended thermospheres give rise to large transit depths in UV transmission spectra.  However, they are also potentially detectable in optical and infrared spectra.  For instance, \citet{coustenis97,coustenis98} searched for an exosphere around 51 Peg b in the near-IR.  In line with the current understanding, they suggested that the exospheres of close-in EGPs such as 51 Peg b are hot and composed primarily of atoms and ions.  They also argued that hydrodynamic escape can lead to the escape of heavier species, and give rise to large in-transit absorption by such species in optical and near-IR spectra.  However, 51 Peg b is not a transiting planet and the search for an exosphere around it was not successful.  On the other hand, once the transit of HD209458b was first detected \citep{charbonneau00,henry00}, similar searches on this planet were also undertaken.  

\citet{moutou01} looked for visible absorption by species such as Na, H, He, CH$^+$, CO$^+$, N$_2^+$, and H$_2$O$^+$ in the upper atmosphere of HD209458b.  These observations were followed by \citet{moutou03} who attempted to measure the transit depth in the He 1083 nm line that was predicted to be significant by \citet{seager00}.  The most recent searches were reported by \citet{winn04} and \citet{narita05} who looked for transits in the visible Na D, Li, H$\alpha$, H$\beta$, H$\gamma$, Fe, and Ca absorption lines.  So far none of the ground-based searches have led to a detection of the upper atmosphere.  However, the non-detection is based on only a few observations that have proven difficult to analyze, and the search should continue.  

Visible and infrared observations have been able to probe the atmosphere of HD209458b at lower altitudes.  In fact, HD209458b was the first EGP to have its atmosphere detected by transmission spectroscopy.  The first detection was achieved by \citet{charbonneau02} who observed a deeper in-transit absorption in the Na D 589.3 nm resonance doublet compared to the adjacent wavelength bands.  This detection was based on four transits observed with the Space Telescope Imaging Spectrograph (STIS) onboard the Hubble Space Telescope (HST) \citep{brown01}.  The same data were later reanalyzed by \citet{sing08a,sing08b} who combined them with other observations \citep{knutson07} and created a transmission spectrum of HD209458b at wavelengths of 300--800 nm.  They argued that the abundance of sodium in the atmosphere is depleted above the 3 mbar level either by condensation into Na$_2$S clouds or ionization.  We note that the detection of Si$^{2+}$ in the thermosphere \citep{linsky10} constrains cloud formation mechanisms in the upper atmosphere and implies that the depletion of Na at low pressures is probably due to ionization (see Section~\ref{subsc:clouds}).                  

The atmosphere of HD209458b has also been observed several times in the infrared with the Spitzer space telescope.  \citet{deming05} detected the secondary eclipse of the planet at 24 $\mu$m by using the Multiband Imaging Photometer (MIPS).  Together with similar observations of TrES-1 at 4.5 and 8.0 $\mu$m obtained by \citet{charbonneau05} who used the Infrared Array Camera (IRAC), these observations constitute the first detections of infrared emission from extrasolar planets.  They were followed by \citet{richardson07} who observed the infrared emission spectrum of HD209458b between 7.5 and 13.2 $\mu$m with the Infrared Spectrograph (IRS).  This spectrum was reanalyzed by \citet{swain08} who noted that it is largely featureless with some evidence for an unidentified spectral feature between 7.5 and 8.5 $\mu$m.  

\citet{knutson08} used IRAC to observe the secondary eclipse of HD209458b in the 3.6, 4.5, 5.8, and 8.0 $\mu$m bands.  They observed a higher than expected flux in the 4.5 and 5.8 $\mu$m bands and interpreted this as evidence for the presence of a stratospheric temperature inversion that gives rise to strong water emission at these wavelengths.  \citet{beaulieu10} observed the transit of the planet in the same wavelength bands and also found evidence for the presence of water vapor in the atmosphere.  The detection of water vapor is also supported by \citet{swain09} who used the Near Infrared Camera and Multi-Object Spectrometer (NICMOS) on HST to observe the secondary eclipse of HD209458b between 1.5 and 2.5 $\mu$m.  In addition to water vapor, the NICMOS observations revealed the presence of methane and carbon dioxide in the emission spectrum.  These detections provide valuable clues to the overall composition of the atmosphere but the uncertainties in the data and degeneracies between temperature and abundances in the forward model prevent a more quantitative characterization of the density and temperature profiles.  

In general, difficulties associated with reducing the data and the need to describe a few uncertain data points with models of growing complexity has led to disagreements on the analysis and interpretation of transmission and secondary eclipse data on different exoplanets.  The same is true of the FUV transit observations of HD209458b.  \citet{vidalmadjar03} used the STIS G140M medium resolution grating to detect a 15~$\pm$~4 \% transit depth in the wings of the stellar H Lyman~$\alpha$ emission line\footnote{The core of the H Lyman~$\alpha$ line is entirely absorbed by the interstellar medium (ISM)}.  Based on the data, they argued that the planet is followed by a cometary tail of escaping hydrogen atoms that are accelerated to velocities in excess of 100 km s$^{-1}$ by stellar radiation pressure \citep[see also][]{schneider98}.  Later, \citet{vidalmadjar04} used the STIS G140L low resolution grating to detect absorption by H, O and C$^+$ in the upper atmosphere, and argued that the atmosphere of HD209458b escapes hydrodynamically.  \citet{linsky10} used the Cosmic Origins Spectrograph (COS) on HST to confirm the detection of C$^+$ and reported on the detection of Si$^{2+}$ around the planet.  They also argued that the atmosphere escapes hydrodynamically.    

\citet{benjaffel07,benjaffel08} disagreed with the interpretation of the G140M observations.  He reanalyzed the G140M data and argued that the 15 \% H Lyman~$\alpha$ transit depth was exaggerated because the data were partly corrupted by short-term variability of the host star and geocoronal Lyman $\alpha$ emissions.  He also showed that there is no evidence for a cometary tail in the data, and that absorption by H in the extended thermosphere of the planet can explain the observations.  \citet{benjaffel10} reanalyzed the G140L data and reached a similar conclusion regarding H \citep[see also][]{koskinen10}.  However, \citet{benjaffel10} fitted the H Lyman~$\alpha$ observations by scaling the density profiles from the model of \citet{garciamunoz07}, and argued that suprathermal O and C$^+$ are required to fit the transit depths in the O I and C II lines.  They did not explain how the suprathermal atoms form and simply chose their properties to fit the observations.  

\citet{holstrom08} offered yet another explanation for the H Lyman~$\alpha$ observations.  They argued that hydrogen atoms cannot be sufficiently accelerated by stellar radiation pressure before they are ionized by stellar X-rays and EUV (XUV) radiation.  Instead, they suggested that the observed absorption arises from a cloud of energetic neutral atoms (ENAs) that form by charge exchange between the protons in the stellar wind and the escaping hydrogen.  In their model, the observed absorption reflects the velocity of the stellar wind, and the data can potentially be used to characterize the magnetosphere of the planet.  \citet{ekenback10} recently updated the model to include a more realistic description of the escaping atmosphere and stellar wind properties.  However, both studies ignored absorption by H in the thermosphere, which is significant, and did not attempt to explain the presence of heavier atoms and ions such as O, C$^+$, and Si$^{2+}$ in the escaping atmosphere.   

All of the interpretations of the UV transit data require that HD209458b has a hot and extended thermosphere.  In fact, \citet{koskinen10} (hereafter K10) also showed that absorption by thermal H and O in such a thermosphere explains both the H Lyman~$\alpha$ and O I 1304 \AA~transit depths.  Further, their model related the observations to a few physically motivated characteristics such as the mean temperature and composition of the upper atmosphere.  They also used the observations to constrain the pressure level where H$_2$ dissociates and estimated that the H$_2$/H transition should occur at 0.1--1 $\mu$bar.  However, these results are based on an empirical model that was simply designed to fit the observations.  One of the aims of the current paper is to show that the results are also supported by more complex physical models.

We have also attempted to establish a more comprehensive description of the thermosphere that treats it as an integral part of the whole atmosphere rather than a separate entity.  In order to do so, we developed a new one-dimensional escape model for the upper atmosphere of HD209458b that includes the photochemistry of heavy atoms and ions, and a more realistic description of heating efficiencies.  This model is described in detail by \citet{koskinen12a} (hereafter, Paper I).  We also used results from a state-of-the-art photochemical model (Lavvas et al., \textit{in preparation}) to constrain the density profiles of the observed species in the lower atmosphere.  We discuss the implications of our results in the context of different observations, and show that observations of the upper atmosphere also yield valuable constraints on the properties of the lower atmosphere.                          
          
\section{Methods}
\label{sc:methods}     

\subsection{Stellar emission lines}
\label{subsc:empirical}

The interpretation of the FUV transit measurements relies on accurate characterization of the stellar emission lines and, provided that the ISM is optically thick over parts of the line profile, their absorption by the ISM.  The observed transit depths also depend on stellar variability.  In this section we discuss the properties of the line profiles, stellar variability, and absorption by the ISM.  The properties of the H Lyman~$\alpha$ and the O I 1304~\AA~triplet lines (hereafter, the O I lines) have been discussed in detail before \citep[e.g.,][]{benjaffel10,koskinen10}, and we present only a brief summary of them here.  However, the COS observations of the C II 1335~\AA~multiplet (hereafter, the C II lines), and the Si III 1206.5~\AA~line (hereafter, the Si III line) \citep{linsky10} have not been modeled before, and thus we constructed emission line models for these lines based on the new data.  These, and the previous models for the H Lyman~$\alpha$ and O I lines, were used to calculate the predicted transit depths in Section~\ref{sc:results}.  The observed line-integrated transit depths for HD209458b are listed in Table~\ref{table:models}.     

\subsubsection{H Lyman~$\alpha$ and O I 1304 \AA~lines}

\citet{wood05} measured the H Lyman~$\alpha$ emission line of HD209458 by using the high resolution echelle E140M grating on STIS, and used the details of the line profile to constrain the column density of H in the ISM along the line of sight (LOS) to HD209458.  Following K10, we adopted the reconstructed line profile and ISM fit parameters from \citet{wood05} for the transit depth calculations in this paper.  The short-term variability of the H Lyman~$\alpha$ emissions from HD209458 was estimated by \citet{benjaffel07} who used the G140M data to derive a magnitude of $\sim$8.6~$\pm$~5.6~\% for this variability -- although the large uncertainty of the observations makes it difficult to separate variability from noise.  There are no observations to constrain the long-term variability or center-to-limb variations of the Lyman~$\alpha$ line on HD209458.  However, typical solar characteristics provide some guidance on these properties.  

\citet{woods00} studied the variability of solar Lyman~$\alpha$ emissions based on satellite observations spanning four and a half solar cycles between 1947 and 1999.  They found that the variability ranges between 1 and 37 \% during one period of solar rotation (27 days), and the average variability during one solar rotation was found to be 9~$\pm$~6~\%.  This result agrees well with the estimated variability of the Lyman~$\alpha$ emissions from HD209458.  The rotation period of HD209458 is estimated to be $\sim$10--11 days \citep{silvavalio08}.  The G140M observations covered three different transits and took place within a month and a half.  Each observation covered approximately 2 hours in time.  Thus the data can be affected by short-term variability and it is essential that such variability be properly accounted for.  For this reason, we compare our models with the results of \citet{benjaffel07,benjaffel08} and \citet{benjaffel10} who analyzed the data in the time tag mode and accounted for variability before calculating transit depths.

Short-term variability in the H Lyman~$\alpha$ line is mostly related to plage activity that is modulated by solar rotation while long-term variability depends on variations in both plage and active network regions \citep{woods00}.  In fact, it is misleading to imagine the transits in any of the FUV emission lines as the planet crossing a smooth, uniformly emitting disk.  Rather, one has to imagine the planet crossing a relatively dark disk with scattered bright regions.  For instance, during solar maximum the plages\footnote{We include enhanced network in the definition of a plage here.} cover approximately 23 \% of the solar disk while active network covers about 8.5 \%.  The brightness contrasts of the plages and active network are 6.7 and 3.2, respectively, when compared to the quiet sun \citep{worden98,woods00}.  This means that the transit depth in the H Lyman~$\alpha$ can vary by factors of 0.4--3 as a function of time during maximum activity.  Similar variability can be expected in the other FUV emission lines.  

The transit depth depends mostly on the path of the planet across the stellar disk although the plage and active network coverage can also change slightly with stellar rotation during the transit.  This highlights the importance of light curve analysis to work out transit depths -- limited `snapshots' during transit may be corrupted by the planet either covering or not covering an active region.  However, it is also possible for the transit depth to be altered even if the variations are not immediately evident in the light curves.  This complicates the analysis of the observations, and underlines the need for repeated observations at different times.  We note that the plage and active network coverage is significantly smaller during solar minimum than it is during solar maximum.  This means that the transit depth is likely to be more stable and closer to the true transit depth during stellar minimum.  Unfortunately, the activity cycle of HD209458 has not been studied in detail, and thus we have no information about it.       

As we noted above, the properties of the O I lines were discussed extensively by K10 who fitted parameterized solar line profiles \citep{gladstone92} to the O I lines of HD209458 that were observed with the STIS E140M grating \citep{vidalmadjar04}.  We adopted the O I line profile and ISM parameters from K10 for this study, and do not discuss these properties further here.  We note that the plage and active network contrasts of the O I lines are 4.2 and 1.7, respectively \citep{woods00}.  These values are slightly smaller than the corresponding values for H Lyman~$\alpha$.  We note that limb darkening or brightening can also affect the observed transit depths \citep[K10,][]{schlawin10}.  However, center-to-limb variations are not particularly significant in the solar H Lyman~$\alpha$ and O I lines \citep{curdt08,rousseldupre85}.   

\subsubsection{C II 1335~\AA~and Si III 1206.5~\AA~lines}

\citet{linsky10} used the medium resolution ($R \sim$~17,500) G130M grating of the COS instrument to observe four transits of HD209458b between September 19 and October 18, 2009 at wavelengths of 1140--1450~\AA.  During each HST visit, they observed the star during transit, at first quadrature, secondary eclipse and second quadrature.  In order to obtain the out-of-transit reference spectra, they co-added the secondary eclipse and quadrature observations from all four visits.  They also co-added the in-transit spectra from all visits to create a single spectrum.  The detections of the transits in the C II 1335 \AA~and Si III 1206.5 \AA~lines were compared with observations of other lines such as Si IV 1395 \AA~in which the transit was not detected.  We note that the transit in the Si III line was not detected earlier by \citet{vidalmadjar04}.  However, these authors use a wider wavelength interval to calculate the line-integrated transit depth and yet report a 2$\sigma$ upper limit of 5.9 \%, implying that the 3$\sigma$ upper limit includes the transit depth observed by \citet{linsky10}.  As we have seen, stellar variability can also cause changes in the perceived transit depth.  Therefore we are not convinced of the reality of a non-detection in the earlier STIS G140L observations.

\begin{figure}
  \centering
  \includegraphics[width=0.7\textwidth]{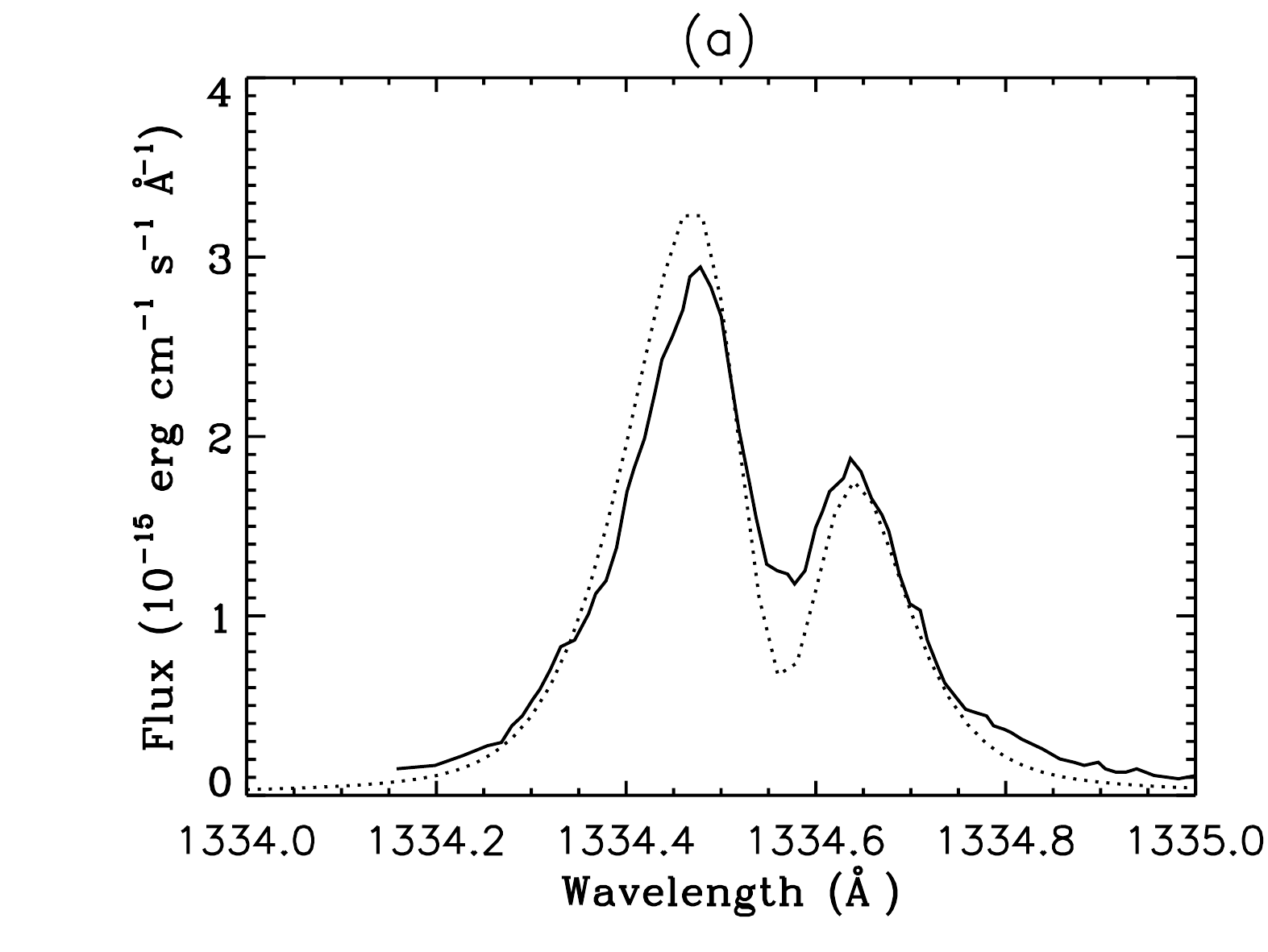}
  \includegraphics[width=0.7\textwidth]{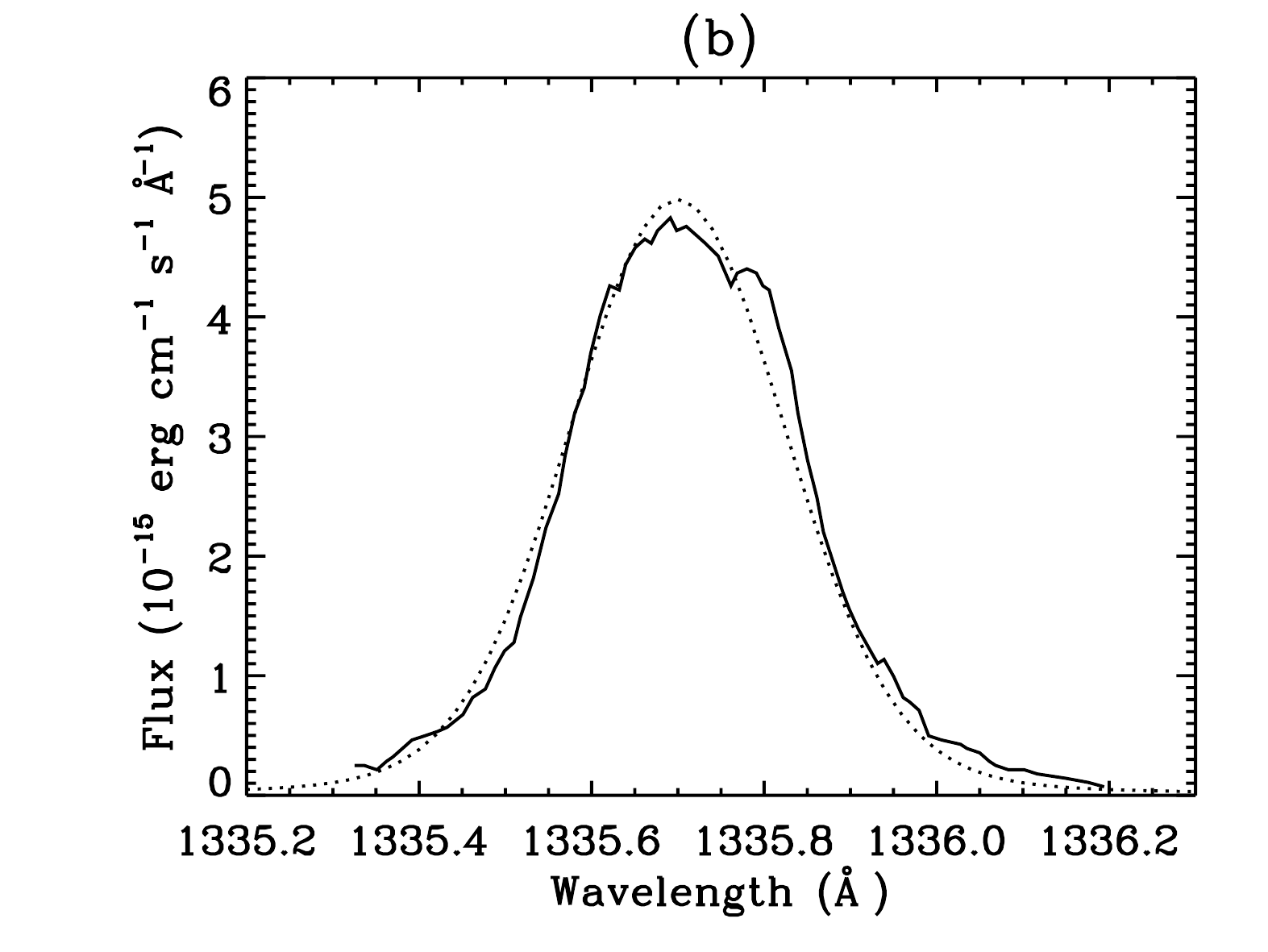}
  \caption{(a) The C II 1334.5 \AA~line of HD209458 \citep[solid line,][]{linsky10} fitted with a Voigt profile (see Table~\ref{table:lines}) and adjusted for absorption by the ISM (dotted line).  We assumed that the column density of ground state C$^+$ in the ISM is 2.23~$\times$~10$^{19}$ m$^{-2}$.  The relative velocity of the ISM with respect to Earth is -6.6 km s$^{-1}$ and the effective thermal velocity along the LOS to the star is 12.3 km~s$^{-1}$ \citep{wood05}.  (b) The C II 1335.7 \AA~line of HD209458 fitted with a Voigt profile.  Absorption by the ISM was assumed to be negligible.  The model profiles were convolved to a spectral resolution of $R =$~17,500.}
  \label{fig:cii}
\end{figure}
                 
In order to calculate transit depths in the C II and Si III lines, we created models for the stellar emission line profiles.  We note that \citet{benjaffel10} also modeled the C II lines in order to fit the transit depth in the low resolution G140L data \citep{vidalmadjar04} where the main components of the C II multiplet are unresolved.  Their model line profiles were also constrained by the high resolution STIS E140M observations of \citet{vidalmadjar03} that resolve the components, although this data has low S/N and it was not used for transit observations.  Here we use the higher S/N COS observations that also resolve the main components of the C II lines to constrain the line profile models.  

The atomic line parameters for the C II and Si III lines are listed in Table~\ref{table:lines}.  The C II multiplet consists of three separate emission lines.  The two lines at 1335.66 \AA~and 1335.71~\AA~(hereafter, the C II 1335.7 \AA~line) are unresolved in the COS (and most solar) observations.  The core of the C II 1334.5 \AA~line is strongly absorbed by the ISM whereas the C II 1335.7 \AA~line is not similarly affected.  The ground state $^2 P_{1/2,3/2}$ of C$^+$ is split into two fine structure levels.  Interstellar absorption of the C II 1335.7 \AA~emissions depends on the population of the $^2 P_{3/2}$ level in the ISM.  In line with a similar assumption regarding the excited states of O (K10), we assumed that this population is negligible.  To estimate the emission line profile from HD209458, we fitted the solar C II 1335.7 \AA~line from the SUMER spectral atlas \citep{curdt01} with a Voigt function and adjusted the result to agree with observations \citep{linsky10}.  Figure~\ref{fig:cii} shows that the model line profile agrees reasonably well with the observed line profile.  The fit parameters for this and other relevant emission lines are listed in Table~\ref{table:lines}.

\begin{table}[htbp]
 \centering
 \caption{Atomic line parameters$^a$}
 \begin{tabular}{@{} lcccccc @{}}
 \toprule
  Line  & $\lambda_{0}$(\AA) & $f_0$ & $A_0$(s$^{-1}$) & $\Delta \lambda_D$(\AA)$^b$ & $\Gamma$ & F(10$^{-15}$ erg cm$^{-2}$ s$^{-1}$) \\ 
 \midrule
  C II          & 1334.53  & 0.1278   & 2.393~$\times$~10$^{8}$ & 0.12   & 0.266 & 1.16 \\
  C II$^c$ & 1335.66  & 0.01277 & 4.773~$\times$~10$^{7}$ & N/A    & N/A     & N/A   \\
  C II          & 1335.71  & 0.1149   & 2.864~$\times$~10$^{8}$ & 0.16   & 0.224  & 1.66 \\
  Si III        &  1206.5      & 1.669     & 2.550~$\times$~10$^{9}$ & 0.12   & 0.535 &  2.09  \\ 
  \bottomrule
  \end{tabular}
  \caption*{\small{$^a$from \citet{morton91}  \\ 
                    $^b$The line profiles are given by $p(\lambda) = [F / (\Delta \lambda_D \sqrt{\pi})] V(a,u)$ where $V$ is the IDL Voigt function with $a = \Gamma/(4 \pi \Delta \lambda_D)$ and $u = (\lambda - \lambda_0)/\Delta \lambda_D$.  The flux is given at Earth distance (47 parsec). \\
                    $^c$The emission line merges with the C II 1335.71 \AA~line.}}
  \label{table:lines}
\end{table}

Strong absorption by the ISM makes fitting the C II 1334.5 \AA~line more complicated than fitting the C II 1335.7 \AA~line.  Following \citet{benjaffel10}, we estimated that the column density of ground state C$^+$ in the ISM is 2.23~$\times$~10$^{19}$ m$^{-2}$ by scaling the column density of C$^+$ measured along the LOS to Capella \citep{wood97} to the distance of HD209458.  We fitted the solar C II 1334.5 \AA~line from the SUMER spectral atlas with a Voigt profile, and used the estimated column density and the results of \citet{wood05} to calculate absorption by the ISM.  We then varied the total flux within the line profile until the results agreed with the observations of \citet{linsky10}.  As a result, we obtained a pre-ISM flux ratio of [C II 1334.5 \AA]/[C II 1335.7 \AA] $\sim$~0.7, which agrees well with the solar value \citep{curdt01}.  The model and observed line profiles are again shown in Figure~\ref{fig:cii}.  We note that the ISM is optically thick at wavelengths between 1334.54 \AA~and 1335.58 \AA, which correspond to Doppler shifts of 2.2 and 11.2 km~s$^{-1}$, respectively.  However, the observed flux in this region is not zero because of spectral line broadening in the COS instrument.

The C II lines are formed in the upper chromosphere and lower transition region of the solar atmosphere.  Similarly with H Lyman~$\alpha$, the brightest emissions are associated with plage activity \citep[e.g.,][]{athay89}.  The plage and active network contrasts for the C II lines are 5.9 and 1.5, respectively \citep{woods00}.  The large spatial variability makes it difficult to characterize center-to-limb variations.  However, different solar observations point to approximately 40~\% limb brightening in the 1335.7 \AA~line and probably a similar variation in the 1334.5 \AA~line, with the intensity rising steadily at $\mu >$~0.6 \citep{lites78,judge03}.  This brightening effect is due to the broadening of the emission line in the limb.      

\begin{figure}
  \centering
  \includegraphics[width=0.7\textwidth]{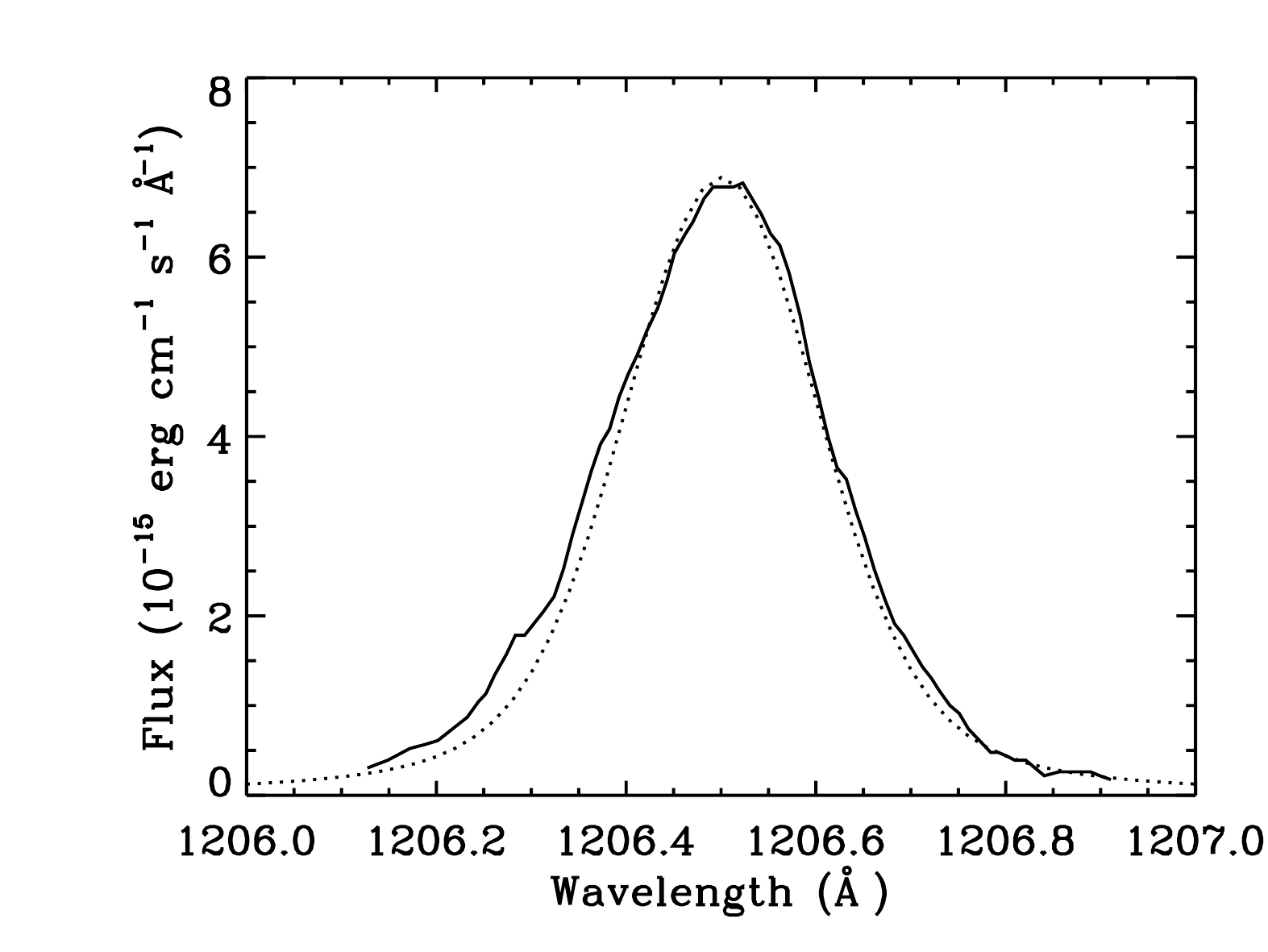}
  \caption{The Si III 1206.5 \AA~line of HD209458 \citep[solid line,][]{linsky10} fitted with a Voigt profile (dotted line, see Table~\ref{table:lines}).  Absorption by the ISM was assumed to be negligible.  The model profile was convolved to a spectral resolution of $R =$~17,500.}
  \label{fig:si3}
\end{figure}

The Si III line arises from a transition between the $^1 S$ ground state and the $^1 P$ excited state.  Again, we fitted the Si III line profile from the SUMER spectral atlas \citep{curdt01} with a Voigt profile and adjusted the resulting line profile to agree with observations of HD209458 \citep{linsky10}.  Figure~\ref{fig:si3} shows the observed and model line profiles, and the fit parameters are listed in Table~\ref{table:lines}.  We assumed that the abundance of Si$^{2+}$ in the ISM is negligible.  This is supported by a lack of detectable absorption by the ISM in the observed line profile.  We note that absorption by the ISM affects the interpretation of the measured transit depths only if parts of the line profile are entirely absorbed or if the properties of the ISM change between observations.  

The solar Si III line has not been studied to the same degree as the C II lines.  However, some constraints on the variability and center-to-limb variations of the Si III line have been obtained.  For instance, \citet{nicolas77} used SKYLAB observations to study the center-to-limb variations of the Si III emissions from the quiet chromosphere.  They found that the line is strongly limb-brightened.  The total line intensity increased by a factor of 2.4 between the disk center and $\mu =$~0.73, and then again by a factor of 3 towards the edge of the limb.  This means that the total intensity in the limb is a factor of $\sim$7 higher than at the disk center, and the line profile is also significantly broader at $\mu >$~0.73 with a self-reversal that does not appear at the disk center.  The Si III emissions from the Sun also exhibit strong spatial and temporal variability \citep[e.g.,][]{nicolas82}.  Limb brightening makes the transit depth appear smaller when the planet is covering the stellar disk while a steeper transit is seen during ingress and egress \citep[e.g.,][]{schlawin10}.  On the other hand, if the planet covers active regions on the disk, the transit depth can appear significantly deeper (K10).  

Spatial variability and limb brightening of the emission lines on HD209458 can potentially be studied through a careful analysis of the transit light curves.  Ideally, the observations should be analyzed in the time tag mode \citep[e.g.,][]{benjaffel07} to identify variations.  This type of reanalysis of the COS data is beyond the scope of this paper.  Instead, we use idealized transit depths based on a uniformly emitting stellar disk in Section~\ref{sc:results} to show that the optical depth of the extended thermosphere in the FUV lines is significant and that the transit in the O I, C II, and Si III lines is comparable to the transit in the H I line.  We consider this sufficient for the present purposes.  

\subsection{Empirical model}
\label{subsc:empirical_model}

K10 developed an empirical model to fit the UV transit observations of HD209458b and other extrasolar planets.  They argued that the H I transits can be explained in terms of three simple parameters that describe the distribution of H in the thermosphere.  These parameters are the pressure $p_0$ where H$_2$ dissociates (the bottom of the H layer), the mean temperature $\overline{T}$ within the H layer and an upper \textit{cutoff level} $r_{\infty}$ based on the ionization of H.  Transits in other emission lines can be explained by fitting the abundance of the heavier absorbers with respect to H.  The details of this model are discussed by K10 and are not repeated here.  Basically it calculates the transit depth observed at Earth orbit by assuming that the planet and its atmosphere constitute a spherically symmetric obstacle with a density profile in hydrostatic equilibrium up to the sonic point (when a sonic point exists).  Absorption by the ISM and spectral line broadening within the observing instrument (STIS or COS) are taken into account.  In Section~\ref{sc:results}, we compare the transit depths calculated by the empirical model with results based on the hydrodynamic model (see below) to show how the empirical model can be used to fit the data and guide the development of more complex models of the upper atmosphere.        
                   
\subsection{Hydrodynamic model}
\label{subsc:hydromodel}

We developed a one-dimensional, non-hydrostatic escape model for HD209458b to constrain the mean temperature and ionization in the upper atmosphere (Paper I).  Results from this model demonstrate that the empirical model is physically meaningful.  It solves the vertical equations of motion for an escaping atmosphere containing H, H$^+$, He, He$^+$, C, C$^+$, O, O$^+$, N, N$^+$, Si, Si$^+$, Si$^{2+}$, and electrons.  The model includes photoionization, thermal ionization, and charge exchange between atoms and ions.  It calculates the temperature profile based on the average solar X-ray and EUV (XUV) flux and up to date estimates of the photoelectron heating efficiencies \citep[][Paper I]{cecchi09}.  The lower boundary of the model is at 1 $\mu$bar and thus the model does not include molecular chemistry.  This is justified because molecules are dissociated by photochemical reaction networks near the 1 $\mu$bar level \citep[e.g.,][]{garciamunoz07,moses11}.  The upper boundary of the model is typically at 16 R$_p$.  We placed the upper boundary at a relatively high altitude above the region of interest in this study, which is below 5 R$_p$.  However, we do not consider the results to be necessarily accurate above 5 R$_p$ (see Paper I for further details).  
 
\section{Results}
\label{sc:results}   

\subsection{Transit depths}
\label{subsc:transits}

In this section we constrain the temperature and composition of the upper atmosphere of HD209458b through a combined analysis of transit observations in the FUV emission lines.  We also compare results from a hydrodynamic model (see Section~\ref{subsc:hydromodel} and Paper I) with the empirical model of K10, and confirm that the latter can be used to constrain the basic properties of the density profiles in the thermosphere.

\subsubsection{Neutral atoms}
\label{subsc:neutrals}

K10 demonstrated that absorption by hydrogen in the extended thermosphere of HD209458b explains the transits in the H Lyman~$\alpha$ line, and used the observations to constrain the mean temperature and composition of the thermosphere.  They fitted both the line-integrated transit depth based on the low resolution G140L data \citep{vidalmadjar04,benjaffel10}, and the transit depths and light curve based on the medium resolution G140M data \citep{vidalmadjar03,benjaffel07,benjaffel08} (see Figures 5 and 6 of K10 for the results).  The results imply that the lower boundary of the absorbing layer of H is at $p_0 =$~0.1--1 $\mu$bar, the mean temperature within the layer is $\overline{T} =$~8,000--11,000 K, and the upper boundary is at $r_{\infty} =$~2.7 $R_p$.  Recent photochemical calculations imply that H$_2$ dissociates near the 1 $\mu$bar level [e.g., Paper I, \citet{moses11}], and this is also supported by an observational lower limit for the vertical column density of H \citep{france10}.  Hence the mean temperature in the thermosphere of HD209458b is approximately 8,250 K (the M7 model of K10). 

\begin{figure}
  \centering
  \includegraphics[width=0.7\textwidth]{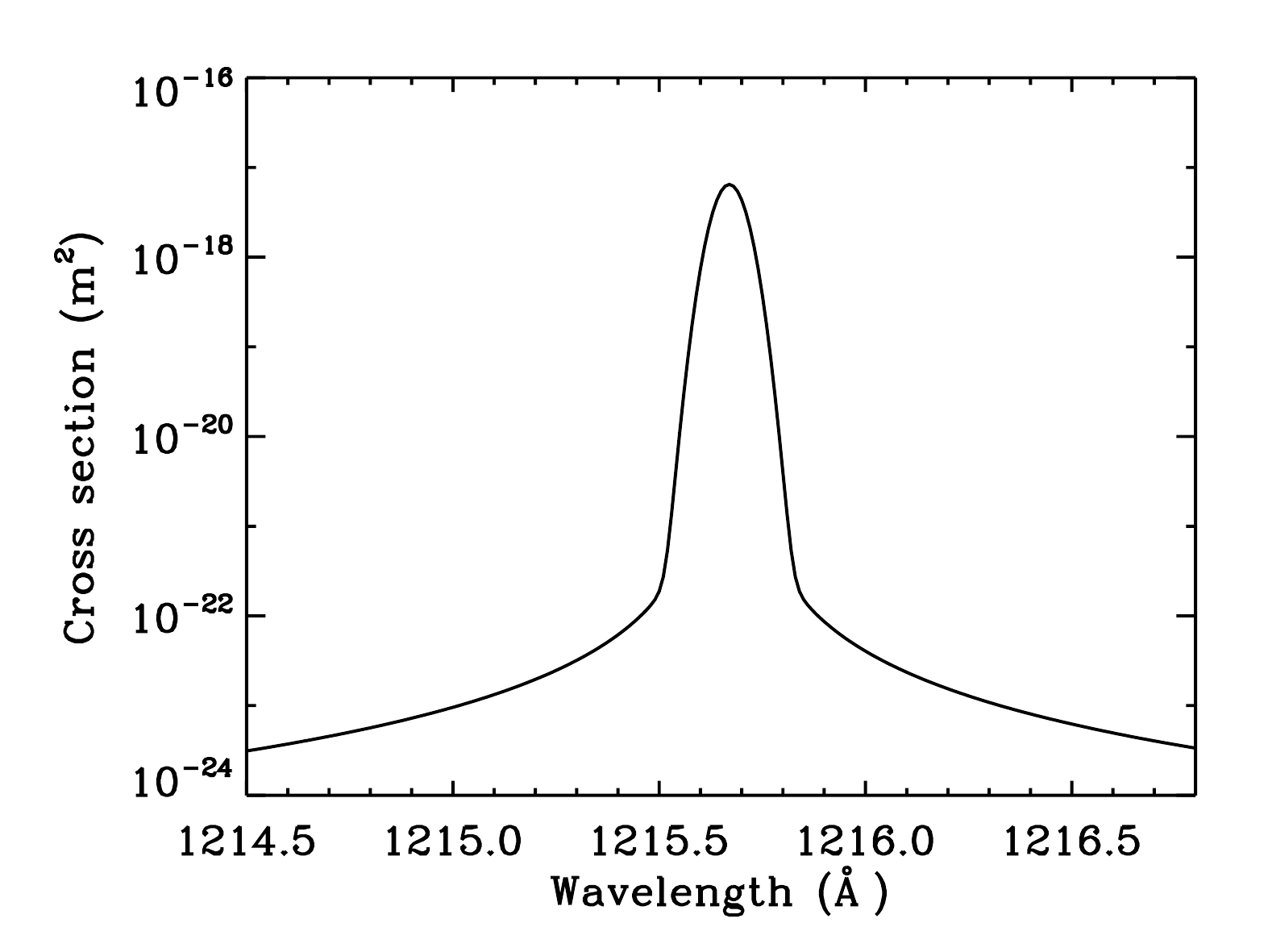}
  \caption{The absorption cross section of H in H Lyman~$\alpha$ line at $T =$~8,250~K.  Thermal and natural line broadening were modeled with a Voigt profile.}
  \label{fig:voigt}
\end{figure}

Here we show from simple arguments that the M7 model agrees in principle with the observed H Lyman~$\alpha$ transit depth.  A similar procedure can also be used to obtain crude estimates of the transit depth for other systems, provided that the properties of the line profile and the ISM are known.  Later in this section we compare the results of the empirical model with results from the hydrodynamic model, and show that they are consistent.  To start with, the line-integrated H Lyman $\alpha$ transit depth of 6.6~$\pm$~2.3~\% \citep{benjaffel10} is consistent with a 6 \% transit depth at 1215.2~\AA~(a Doppler shift of -120 km~s$^{-1}$ from the line core) i.e., the blue peak of the observed line profile (Figure 5 of K10).  Assuming that the extended atmosphere is spherically symmetric, a 6~\% transit depth measured in the blue wing (bw) of the line profile implies an optical depth of $\tau_{\textrm{bw}} \approx$~1 at the impact parameter of 2.1~$R_p$ from the center of the planet.  Figure~\ref{fig:voigt} shows the absorption cross section of H in the Lyman~$\alpha$ line at a temperature of 8,250 K.  The cross section at 1215.2~\AA~is $\sigma_{\textrm{bw}} =$~2~$\times$~10$^{-23}$ m$^2$ and thus a LOS column density of $N_{\textrm{H}} =$~5~$\times$~10$^{22}$ m$^{-2}$ at 2.1 R$_p$ is required to explain the observed absorption. 

\citet{ekenback10} and \citet{lammer11} argued that the optical depth of H in the thermosphere of HD209458b is not significant and instead a large cloud of energetic neutral atoms (ENAs) is required to explain the H Lyman~$\alpha$ observations.  In particular, \citet{lammer11} claimed that a column density of $N_{\textrm{H}} \approx$~10$^{31}$ m$^{-2}$ in the thermosphere is required for strongly visible absorption.  It is easy to see that this estimate is not correct because the wings of the line profile become optically thick with column densities much smaller than this.  It also disagrees with \citet{garciamunoz07} and \citet{benjaffel07,benjaffel08} who were the first to suggest that H in the thermosphere may be sufficiently abundant to explain the observations.  This basic result has also been confirmed by more recent calculations by \citet{trammell11}.  All of these calculations show that the optical depth of the thermosphere below 3 $R_p$ is not negligible.  

The empirical model of K10 is simplified by the use of the hydrostatic approximation.  This is well justified even if the atmosphere is escaping.  In general, the density profile of the escaping gas in the thermosphere can be estimated from \citep{parker64a}:
\begin{equation}
n(\xi) c^2 (\xi) = n_0 c_0^2 \exp \left( -\int_1^{\xi} \frac{\textrm{d}u}{u^2} \frac{W^2}{c^2} \right) \exp \left(-\int_1^{\xi} \frac{du}{c^2} v \frac{dv}{du} \right)
\label{eqn:parkerden}
\end{equation}
where $\xi = r/r_0$, $c^2(\xi) = k T(\xi)/m$, $W = G M_p / r_0$, $m$ is the molecular weight, and $v$ is the vertical flow speed.  For convenience, we retained Parker's original notation.  The first integral on the right hand side of equation~(\ref{eqn:parkerden}) applies in hydrostatic equilibrium.  The second integral is negligible below the sonic point for transonic escape and always negligible for subsonic escape (i.e., evaporation).  The sonic point on HD209458b is always above 3 $R_p$ (Paper I) and thus the density profile of H is approximately hydrostatic at least up to 2.1 $R_p$.  

Assuming a hydrostatic atmosphere, the LOS column density of H at 2.1 $R_p$ can be estimated from:
\begin{equation}
N_{\textrm{H}} (r = 2.1 \  \textrm{R}_p) \approx n_0 \exp \left[ \frac{G M_p m}{k \overline{T}} \left( \frac{1}{r} - \frac{1}{r_0} \right) \right] \sqrt{2 \pi r H( r )}
\label{eqn:hcolumn}
\end{equation}
where $\overline{T}$ and $H$ are the mean temperature and scale height, respectively.  We note that the mean thermal escape parameter is:
\begin{equation}
\overline{X}( r ) = \frac{G M_p m}{k \overline{T} r}.
\end{equation}
Assuming that $\overline{T} =$~8,250 K and using the planetary parameters of HD209458b ($M_p =$~0.7 $M_J$, $R_p =$~1.3 $R_J$), we obtain values of $\overline{X}_0 =$~13.8 and $\overline{X} (2.1 \  \textrm{R}_p) =$ 6.6.  Thus, according to equation~(\ref{eqn:hcolumn}), the column density of $N_{\textrm{H}} =$~5~$\times$10$^{22}$ m$^{-2}$ that is required to explain the observations implies that $n_0 =$~3.5~$\times$~10$^{17}$ m$^{-3}$.  This in turn means that $p_0 =$~0.4 $\mu$bar i.e., close to the lower boundary of the M7 model (here the agreement with the M7 model is obviously not exact because the transit depths in K10 are based on a complete forward model of the observed transit within the whole line profile).

The parameters of the empirical model can be compared with corresponding values derived from the hydrodynamic models presented in Paper I (see Section~\ref{subsc:hydromodel} here for a brief summary).  For instance, the mean temperature of the empirical model corresponds roughly to the pressure-averaged temperature of the thermosphere, which is given by \citep[e.g.,][]{holton04}:
\begin{equation}
\overline{T_p} = \frac{\int_{p_1}^{p_2} T( p ) \  \textrm{d} (\ln p)}{\ln (p_2/p_1)}.
\label{eqn:meantemp}
\end{equation}
The hydrodynamic calculations show that the pressure averaged (mean) temperature below 3 $R_p$ based on the average solar flux varies between 6,000 K and 8,000 K for net heating efficiencies $\eta_{\text{net}}$ between 0.1 and 1.  This temperature is relatively insensitive to different assumptions about heating efficiencies or the upper boundary conditions.  In the reference C2 model of Paper I the mean temperature is 7,200 K.  This model is based on our best estimate of the heating efficiencies that are appropriate in the strongly ionized upper atmosphere of HD209458b.  

With a cutoff level at 2.7 $R_p$, we obtained a line-integrated H Lyman~$\alpha$ transit depth of 4.7 \% based on the density of H in the C2 model.  This value agrees with the observations to within 1$\sigma$ \citep{vidalmadjar04,benjaffel10}, but it is smaller than the transit depth of 6.6 \% predicted by the M7 model.  One reason for this is the lower mean temperature of the C2 model.  In order to facilitate a direct comparison between the hydrodynamic model and the empirical model, we calculated the empirical transit depth based on the mean temperature of 7,200 K (hereafter, the M7b model).  The line-integrated transit depth based on this model is 5.8 \%, which is still higher than the transit depth based on the C2 model.  

\begin{figure}
  \centering
  \includegraphics[width=0.7\textwidth]{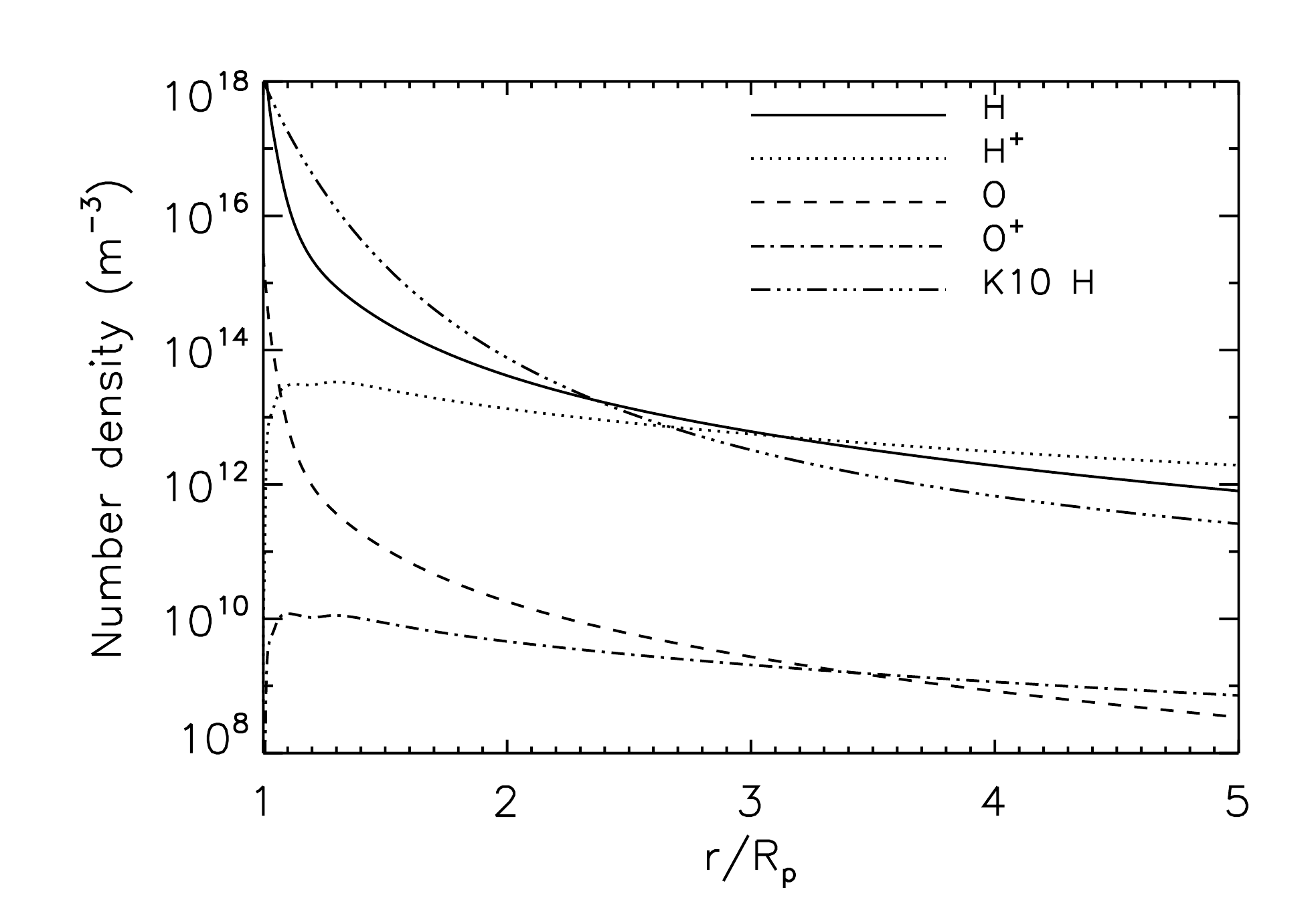}
  \caption{Density profiles O and H based on the C2 model (Paper I) and the density profile of H based on the empirical model of K10 with a mean temperature of 7,200 K.}
  \label{fig:neutrals}
\end{figure}

Figure~\ref{fig:neutrals} shows the density profiles of H, H$^+$, O, and O$^+$ from the C2 model, and the density profile of H from the M7b model.  The difference between the transit depths based on the empirical and hydrodynamic models arises because the C2 model has large temperature gradients (Paper I) and a gradual H/H$^+$ transition rather than a sharp cutoff.  The difference does not arise because the density profile of the C2 model deviates from hydrostatic equilibrium.  Given the temperature gradient in the model, the density profile is almost exactly in hydrostatic equilibrium below 3 $R_p$.  In fact, the neutral density profile of the C2 model is better represented by a mean temperature of 6,300 K (not shown).  This  implies that the correspondence of the mean temperature of the empirical model and the pressure averaged temperature of the hydrodynamic model is relatively good but not exact.  

\begin{figure}
  \centering
  \includegraphics[width=0.7\textwidth]{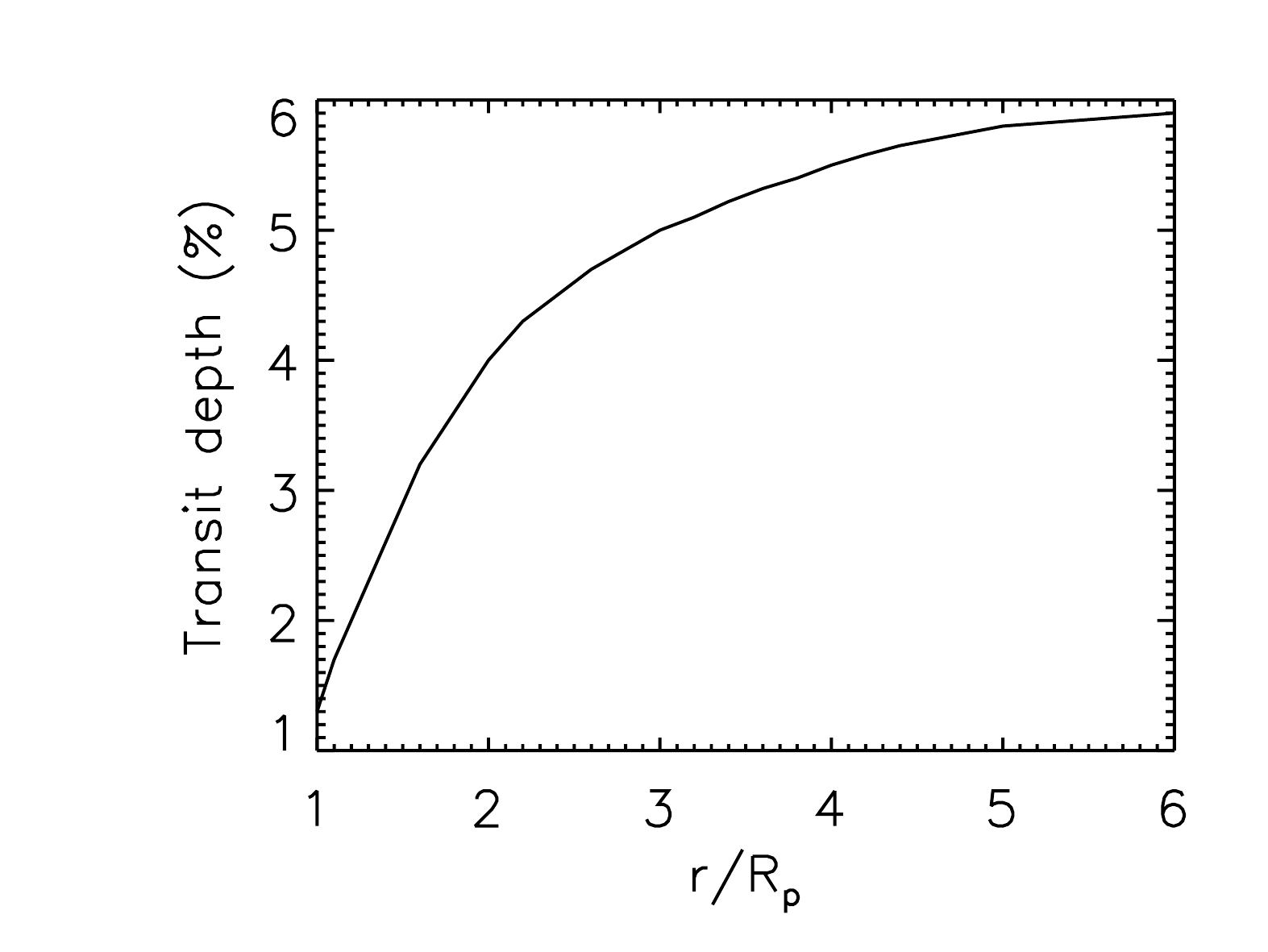}
  \caption{Line-integrated transit depth within the H Lyman~$\alpha$ line as a function of the obstacle cutoff level based on the C2 hydrodynamic model (Paper I).  The K10 density profile applies to the M7b model (see text).}
  \label{fig:cutoff}
\end{figure}

A better agreement between the transit depths based on the C2 and M7b models is obtained with the cutoff level of the C2 model at 5 $R_p$ (see Table~\ref{table:models}).  Figure~\ref{fig:cutoff} shows the line-integrated transit depth within the H Lyman~$\alpha$ line profile as a function of the cutoff level for the C2 model.  This figure indicates that the transit depth increases less steeply with altitude above 4 $R_p$ than it does below this cutoff level, and saturates near 5 $R_p$.  This is a natural consequence of the fact that the LOS column density decreases approximately exponentially with altitude in the lower thermosphere and the wings of the line profile become optically thin at high altitudes.  

We note that K10 already compared the results based on the M7 model with in-transit transmission as a function of wavelength within the H Lyman~$\alpha$ line profile and the light curve derived from the G140M data \citep{benjaffel07,benjaffel08}.  It is not necessary to repeat a similar comparison here because spherically symmetric models that predict line-integrated transit depths that agree with the measured values are generally compatible with both the G140L and G140M data.  This is partly because the uncertainty of the individual data points within the line profile and the light curve is large, and thus they do not strongly constrain the properties of the atmosphere.

According to Figure~\ref{fig:neutrals}, the H/H$^+$ transition in the C2 model occurs near 3.1 $R_p$.  The exact location depends on photochemistry and vertical velocity, and generally the transition occurs near or above 3 $R_p$ (Paper I).  With a fixed pressure at the lower boundary, a faster velocity leads to a transition at a higher altitude.  These results disagree with \citet{yelle04} and \citet{murrayclay09} who predicted a lower transition altitude, but they agree qualitatively with the solar composition model of \citet{garciamunoz07}.  The density profiles of O and O$^+$ are strongly coupled to H and H$^+$ by charge exchange.  As a result, the O/O$^+$ transition occurs generally near the H/H$^+$ transition.  For instance, in the C2 model it is located near 3.4 $R_p$.  We note that significant ionization of H and O above 3 $R_p$ is anticipated by K10 and there is good agreement on this between the empirical and hydrodynamic models.  

The detection of O at high altitudes constrains the mass loss rate and the ionization state of the upper atmosphere (see Section~\ref{subsc:ionescape}).  However, the large uncertainty in the observations means that repeated observations are required to confirm the transit depth.  The M7 model of K10 with a solar O/H ratio \citep{lodders03} yields an O I transit depth of 3.9 \%, which is within 1.5$\sigma$ of the observed value and therefore a satisfactory fit to the data.  This value agrees well with the O I transit depth based on the C2 model (see Table~\ref{table:models}).  K10 argued that a higher transit depth is possible if the mean temperature is higher and/or if the O/H ratio is enhanced with respect to solar.  The hydrodynamic calculations indicate that the latter option is more favorable because higher temperatures lead to stronger ionization of O and may not help to significantly enhance the transit depth.  Indeed, the predicted transit depth agrees with the observations to better than 1$\sigma$ if the O/H ratio is enhanced by a factor of 5 relative to solar (see the MSOL2 model in Table~\ref{table:models}).          

\begin{table}[htbp]
 \tiny{
 \centering
 \caption{Model parameters and transit depths (\%)}
 \begin{tabular}{@{} cccccccccc @{}}
 \toprule
 & & & & H I$^{\textrm{c}}$ & O I & C II 1334.5 \AA & C II 1335.7 \AA & Si III \\
 Model & $\dot{M}$ (10$^7$ kg~s$^{-1}$)  & $\eta_{\text{net}}$$^{\text{a}}$ & $\overline{T}_p$$^{\text{b}}$ (K) & 6.6~$\pm$~2.3$^{\text{d}}$ & 10.5~$\pm$~4.4 & 7.6~$\pm$~2.2 & 7.9~$\pm$~1.5 & 8.2~$\pm$~1.4 \\
 \midrule
  C1         &   5.6   &  0.56  & 7250 &  5.5  &  4.1  &  3.4  &  6.8   &  5.0  \\
  C2         &   4.0   &  0.44  & 7200 &  5.7  &  4.0  &  3.2  &  6.7   &  4.6  \\
  C3         &   6.4   &  0.57  & 6450 &  5.4  &  3.7  &  3.1  &  5.5   &  3.5  \\
  C4         &   4.5   &  0.46  & 7110 &  5.2  &  3.7  & 3.2   &  6.3   &  4.6 \\
  C5         &   5.6   &  0.56  & 7290 &  5.6  &  4.2  &  3.5  &  6.9   &  5.1 \\
  C6         &   3.9   &  0.45  & 7310 &  5.9  &  4.1  &  3.3  &  6.9   &  4.6 \\
  SOL2    &   11.0 &  0.50  & 7390 &  5.0  &  4.2  &  4.3  &  7.8   &  6.8 \\
  MSOL2 &  6.0    &  0.66  & 7370 &  5.6  &  7.1  &  5.5  &  10.5 &  8.0 \\  
  M7         &  N/A   &  N/A   & 8250 &  6.6  &  3.9  &  3.9  &  8.0   &   5.8 \\                                           
  \bottomrule
  \end{tabular} 
  \caption*{\small{$^a$Net heating efficiency (see Section~\ref{subsc:hydromodel}) i.e., the ratio of the net heating flux at all wavelengths to the unattenuated stellar flux (0.45 W~m$^{-2}$) at wavelengths shorter than 912~\AA. \\ 
                    $^b$Pressure averaged temperature below 3 $R_p$. \\
                    $^c$Line-integrated transit depth.  The upper boundary of the absorbing atmosphere is at 5 R$_p$ apart from the M7 model where it is at 2.7 $R_p$ for neutrals and at 5 $R_p$ for ions. \\
                    $^d$Observed values from \citet{benjaffel10} and \citet{linsky10} with 1$\sigma$ errors.}}  
  \label{table:models}}
\end{table}

We have now verified that the empirical model can be used to constrain the mean temperature and extent of the absorbing layer in the thermosphere of HD209458b.  In particular, the comparison of the empirical model with the hydrodynamic model shows that the results of K10 were not affected by the simplifying assumption of hydrostatic equilibrium.  We note that the purpose of the empirical model is to identify physical processes that might otherwise be missed in more complex models that are often based on a large number of uncertain assumptions.  The results from \textit{any} model can now be compared with the observations by identifying the limits of the absorbing layer and calculating the global pressure averaged temperature within that layer.  The values can then be compared with the parameters of the best-fit empirical model.  

\subsubsection{Ions}  
\label{subsc:ions}   

\begin{figure}
  \centering
  \includegraphics[width=0.7\textwidth]{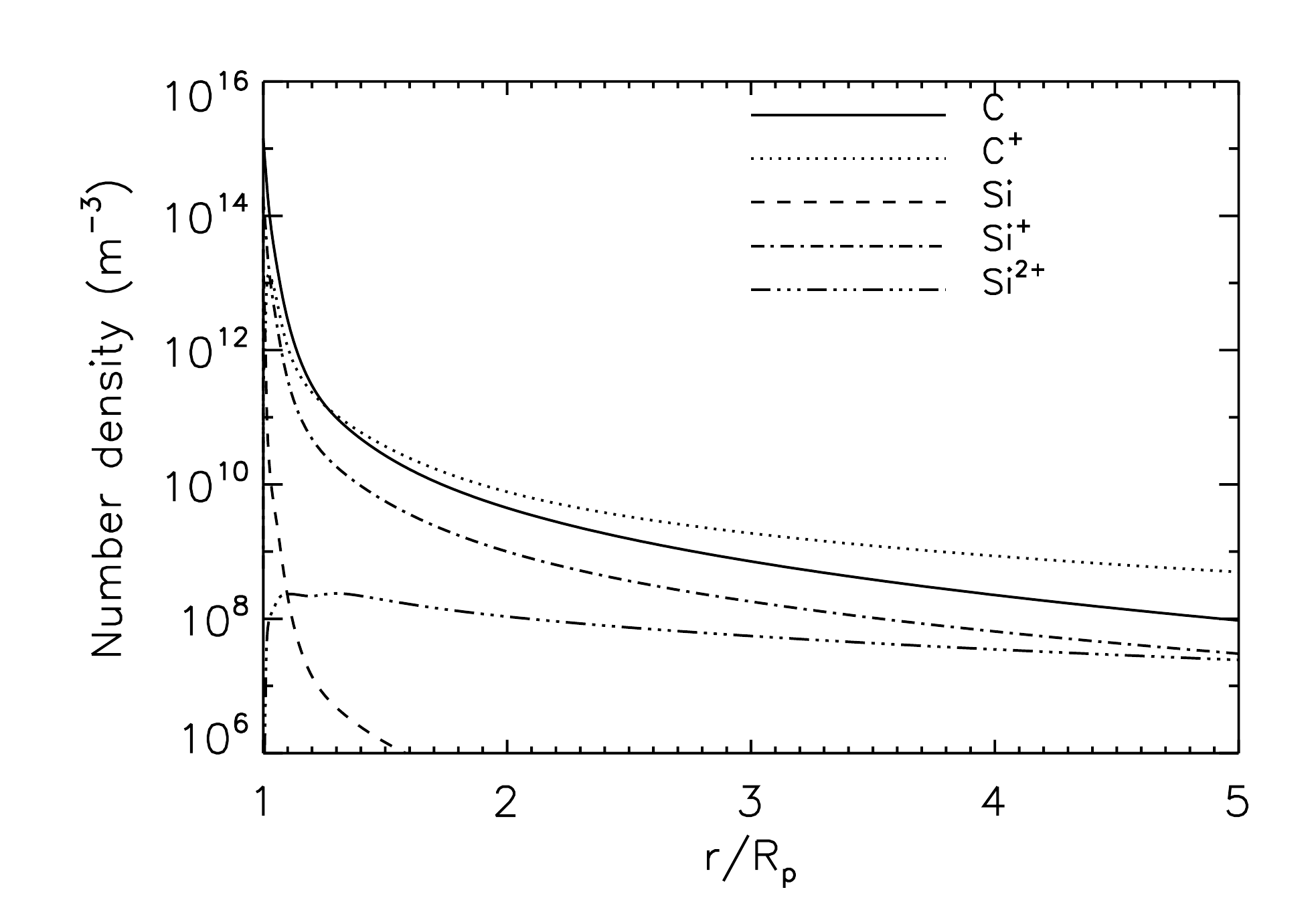}
  \caption{Density profiles of C and Si in the upper atmosphere of HD209458b based on the C2 model (Paper I).}
  \label{fig:ions}
\end{figure}

Figure~\ref{fig:ions} shows the density profiles of the carbon and silicon ions in the thermosphere of HD209458b based on the C2 hydrodynamic model (Paper I).  The major carbon and oxygen-bearing species in the lower atmosphere, CO and H$_2$O, dissociate near the 1 $\mu$bar level \citep[e.g.,][or Lavvas et al., \textit{in preparation}]{moses11}.  Thermochemical calculations indicate that SiO is the dominant silicon-bearing gas on HD209458b \citep{visscher10}.  The detection of Si$^{2+}$ in the upper atmosphere implies that the formation of silicon clouds in the lower atmosphere is suppressed (see Section~\ref{subsc:clouds}), and SiO is also dissociated near 1 $\mu$bar either thermally or by photochemistry.  Thus we assumed that only atomic carbon and silicon are present in the thermosphere, initially with solar abundances \citep{lodders03}.  

According to Figure~\ref{fig:ions}, the C/C$^+$ transition occurs at a much lower altitude of 1.2 $R_p$ than the H/H$^+$ and O/O$^+$ transitions.  Silicon is also almost fully ionized at the lower boundary of the model, and the Si$^{2+}$/Si$^+$ ratio is about 10 \% below 3 $R_p$.  These results are in qualitative agreement with the observations, because they show that H and O are mostly neutral below 3 $R_p$ whereas C and Si are mostly ionized, and a significant abundance of Si$^{2+}$ is possible.  However, it is also useful to explore if the results from the models are in quantitative agreement with the observations and if not, to adjust the model parameters as necessary to explain the data.     

With a cutoff level at 2.7 $R_p$, we obtained line-integrated transit depths of 2.3~\%, 3.6~\%, and 2.5~\% in the C II 1334.5 \AA, C II 1335.7 \AA~and Si III lines, respectively, based on the C2 model.  Here we assumed that the population of the $^2 P$ levels of C$^+$ are in LTE.  We also calculated empirical transit depths based on the M7 model.  In order to do this we assumed that both C and Si are ionized at the lower boundary, and that 10 \% of silicon is Si$^{2+}$.  With these assumptions we obtained C II 1334.5 \AA~and 1335.7 \AA~transit depths of 2.7~\% and 4.2~\%, respectively, and a Si III transit depth of 3 \%.  These values agree well with the C2 model, and further demonstrate the consistency of the empirical and hydrodynamic models.  However, they deviate from the observed values by more than 2$\sigma$.  

The cutoff level of the empirical model is somewhat arbitrary.  For neutral species it is partly based on ionization (K10), but this criterion obviously does not apply to ions.  With a cutoff level at 5 $R_p$ for the ions only, the M7 model yields line-integrated transit depths of 3.9 \%, 8 \%, and 5.8 \% in the C II 1334.5 \AA, C II 1335.7~\AA, and Si III lines, respectively, if 40 \% of silicon is Si$^{2+}$.  These values agree with the observed values to better than 2$\sigma$.  Similarly, by extending the cutoff level of the C2 model to 5 $R_p$, we obtained transit depths of 3.2 \%, 6.7 \%, and 4.6 \% in the C II 1334.5 \AA, C II 1335.7~\AA, and Si III lines, respectively (see Table~\ref{table:models}).  These values deviate from the observed values by 2$\sigma$, 0.9$\sigma$, and 2.6$\sigma$, respectively.  The transit depths predicted by the M7 model are higher partly because the mean temperature of 8,250 K is higher than the corresponding temperature in the C2 model (Paper I).  This also leads to the higher Si$^{2+}$/Si$^+$ ratio that we used here.  

It is not clear if the apparent disagreement between the models and some of the observations needs to be taken seriously.  Stellar activity and other uncertainties mean that the true transit depth can differ from measured values by a significant factor (see Section~\ref{subsc:empirical}).  Further, the C2 model agrees with the line-integrated H Lyman~$\alpha$ and C II 1335.7~\AA~transit depths to within 1$\sigma$, and with the O I and C II 1334.5~\AA~lines to within 2$\sigma$.  Thus we could argue that the present observations are roughly consistent with solar abundances and heating based on the average solar XUV flux.  Nevertheless, we explore the apparent disagreement between the C2 model and the observations further below.  This disagreement is limited to the O I, C II 1334.5~\AA, and Si III lines.  

\begin{figure}
  \centering
  \includegraphics[width=0.7\textwidth]{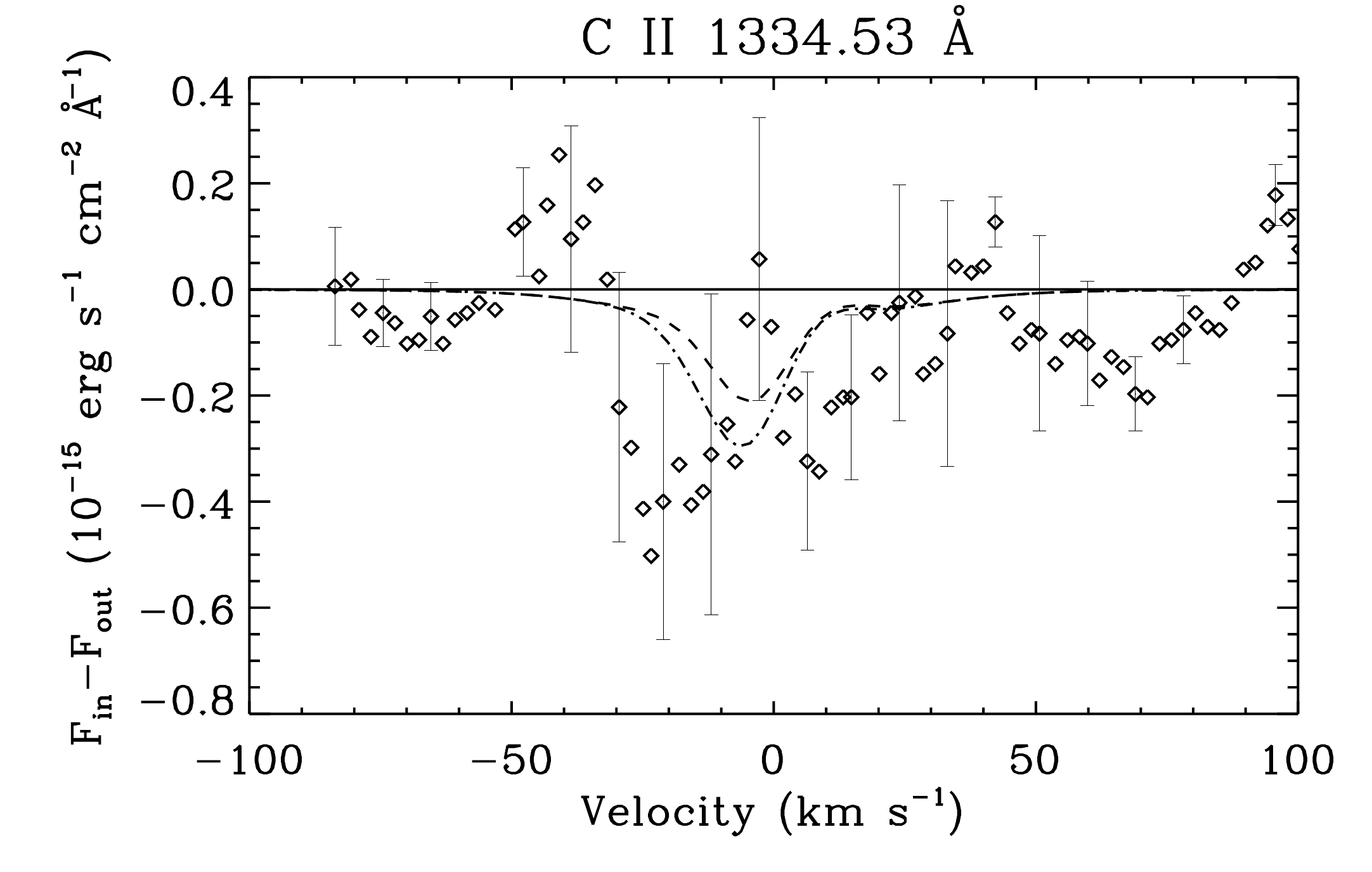}
  \includegraphics[width=0.7\textwidth]{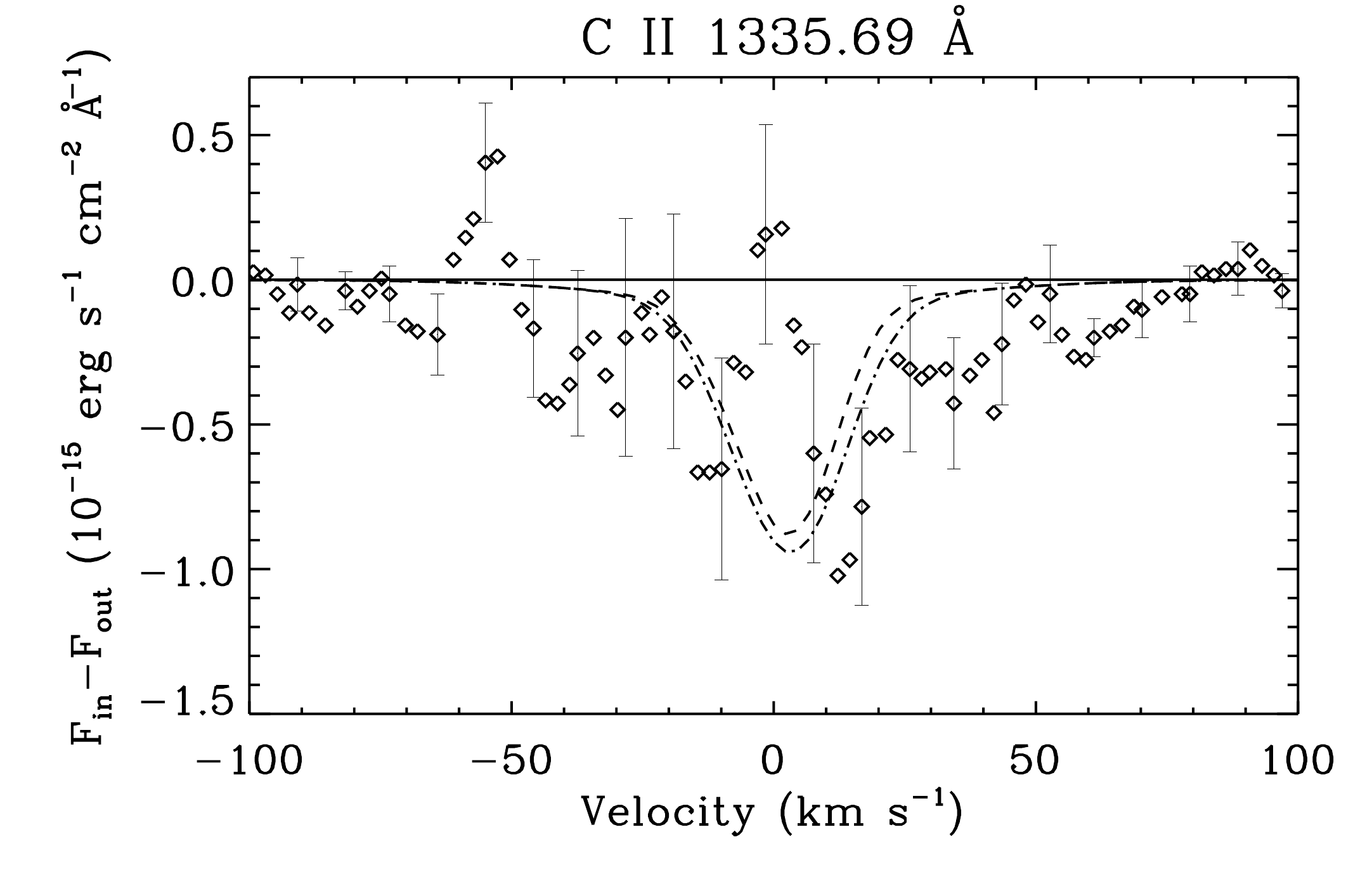}
  \includegraphics[width=0.7\textwidth]{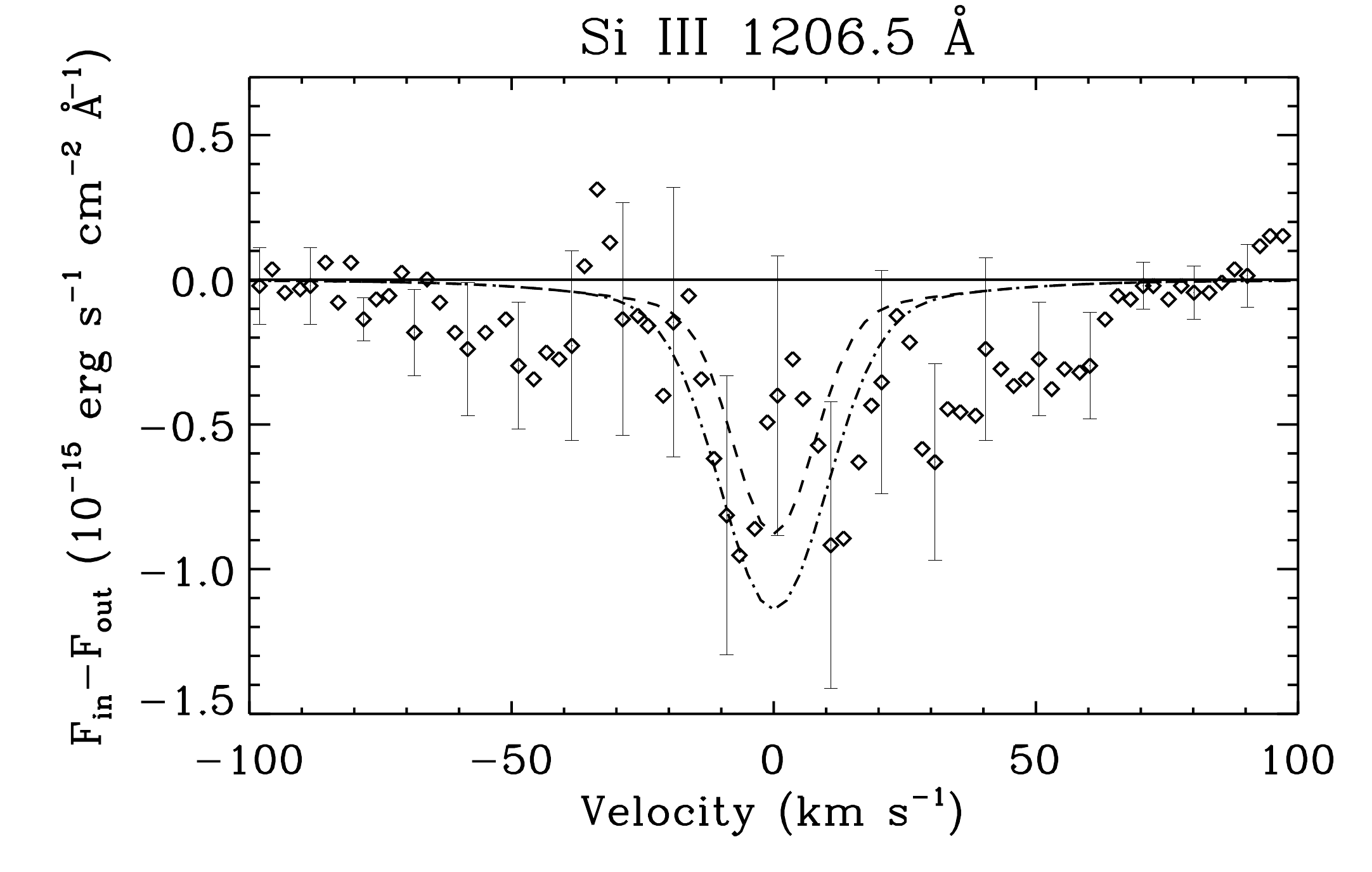}
  \caption{Flux differences between the stellar in-transit and out-of-transit C II and Si III emission lines.  The data points were taken from \citet{linsky10}.  Also shown are model predictions based on the C2 model (dashed line) and the SOL2 model (dash-dotted line).  See text and Table~\ref{table:models} for the details of these models.}
  \label{fig:transits}
\end{figure}

Figure~\ref{fig:transits} shows the observed in-transit flux differences in the C II and Si III lines as a function of wavelength together with different model predictions.  The observations indicate that the transit depths based on the C2 model fall short of the observed values because the model underestimates the width of the absorption lines.  \citet{linsky10} argued that there is velocity structure within the line profiles near Doppler shifts of -10 km~s$^{-1}$ and 15 km~s$^{-1}$ that accounts for broad absorption.  However, the uncertainty of the individual data points is too large to constrain the shape of the absorption lines in detail.  We agree with \citet{linsky10} that the presence of velocity structure needs to be confirmed by future observations.  Thus the additional absorption could also arise from spectral line broadening. 

We note that H$_2$ absorbs within the Si III line and it can contribute to the observed absorption.  The cross section has a lot of structure in this wavelength region, but for the sake of the argument we adopted a mean cross section of $\sim$2~$\times$~10$^{-23}$ m$^{-2}$ \citep{backx76}.  A transit depth of 8 \% across the line profile implies that the atmosphere is optically thick up to 2.4 $R_p$ where the LOS optical thickness is $\tau \sim$~1, and thus a H$_2$ column density of 5~$\times$~10$^{22}$ m$^{-2}$ along this LOS would be required to explain the observed transit depth.  This is unrealistic because the required column density is higher than the corresponding column density of H (see Section~\ref{subsc:neutrals}).  Also, according to our photochemical calculations H$_2$ dissociates in the thermosphere, and the mixing ratio of H$_2$ falls below 0.1 above the 0.1 $\mu$bar level.  Such a low abundance of H$_2$ has no effect on the Si III transit depth.

A higher mean temperature leads to higher transit depths.  Therefore we generated a new hydrodynamic model by multiplying the average solar flux by a factor of 2 and assumed a net heating efficiency of $\eta_{\text{net}} =$~0.5 (hereafter, the SOL2 model).  This model agrees to better than 1$\sigma$ with all of the observed transit depths apart from the O I and C II 1334.5~\AA~lines (see Figure~\ref{fig:transits} and Table~\ref{table:models}).  The mean temperature of the model below 3 $R_p$ is 7,400 K, which is lower than the mean temperature in the M7 model.  However, absorption by the SOL2 model is strengthened by velocity dispersion within the escaping plasma that is not included in the M7 model.  In general, the outflow velocity of the SOL2 model is significantly higher than the velocity in the C2 model.  The model also predicts a mass loss rate of 10$^8$ kg~s$^{-1}$, which is twice as high as the mass loss rate based on the C2 model.  This proves that a higher stellar XUV flux or a corresponding alternative energy source can explain the observations.  An enhancement of the average solar flux by a factor of 2 is not unreasonable, and would roughly correspond to solar maximum conditions.

In Paper I we noted that the stellar XUV flux, or the corresponding alternative heat source, would have to be 5--10 stronger than the average solar flux to produce a mean temperature between 8,000--9,000 K.  Under such circumstances, the predicted transit depths in the C II and Si III lines would obviously be even higher than the values predicted by the SOL2 model.  Indeed, higher temperatures broaden absorption in the wings of the line profiles and may help to explain the in-transit flux differences better (see Figure~\ref{fig:transits}).  However, the energy input and temperature in the model cannot be increased without bound.  Higher temperatures and flux lead to more efficient ionization of the neutral species, and as a result the transit depths in the H Lyman~$\alpha$ and O I lines begin to decrease.  Also, mass loss rates of 10$^{9}$--10$^{10}$ kg~s$^{-1}$ lead to the loss of 10--100 \% of the planet's mass over the estimated lifetime of the system, and this probably limits reasonable energy inputs to less than $\sim$10 times the solar average on HD209458b.

In addition to higher temperature and velocity, supersolar abundances of O, C, and Si can also lead to higher transit depths.  This option is interesting because it also allows for a higher transit depth in the O I lines.  As an example, we generated the MSOL2 model by enhancing the solar O/H, C/H, and Si/H abundances in the hydrodynamic model by a factor of 5.  As a result, we obtained transit depths that agree with nearly all of the observed line-integrated transit depths to better than 1$\sigma$ (see Table~\ref{table:models}).  However, the MSOL2 model overestimates the line-integrated C II 1335.7~\AA~transit depth, and generally overestimates absorption within the cores of the C II and Si III lines.  This could imply that a higher temperature or some other source of additional broadening is a better explanation of the Si III and C II transit depths while a supersolar O/H ratio is required to match the measured O I transit depth.  However, the data points in Figure~\ref{fig:transits} and the observed O I transit depth are too uncertain and do not allow for firm constraints on this.  

Enrichment of heavy elements is a common feature on the gas and ice giants in the solar system.  For instance, the C/H, N/H, S/H, Ar/H, Kr/H, and Xe/H ratios in the atmosphere of Jupiter are all enriched by factors of 2--3 with respect to solar abundances \citep[e.g.,][]{mahaffy00,wong04}.  On Saturn, on the other hand, the C/H ratio is enriched by a factor of 10 \citep{flasar05,fletcher09}.  Enrichment by factors of 4--20 is expected in the N/H and S/H ratios, although condensation of NH$_3$ and H$_2$S in the deep atmosphere of Saturn makes it difficult to constrain the abundances precisely \citep[see][for a review]{fouchet09}.  On Neptune and Uranus the C/H ratio is believed to be enriched by factors of 30--50 \citep[e.g.,][]{owen03,guillot07}, and similar enrichments are possible in the abundance ratios of some of the other heavy elements.  Substantial enrichment of heavy elements with respect to solar abundances is therefore also feasible in EGP systems even if the metallicity of the star is close to solar.   

Unfortunately, we cannot use the current observations to constrain the elemental abundances of the atmosphere with accuracy.  In this regard, the large uncertainty of the observations is unfortunate, because similar observations can potentially be used to estimate them.  The dissociation of molecules at the relatively high pressure of 1 $\mu$bar and the lack of diffusive separation mean that the abundances of the heavy atoms and ions are simply dependent on the elemental abundances and ionization rates.  Observations of the neutral species can therefore be used to constrain the temperature and ionization state, and thus the elemental abundances of the heavy species, but the S/N of the current data does not allow for strong constraints.    

It is interesting to note that while the velocity structure of the escaping plasma can lead to broader absorption that helps to explain the transit depths, it is not necessarily detectable in the data.  For instance, Figure~\ref{fig:transits} shows the transit depths based on the SOL2 model that has a relatively high radial velocity reaching 11 km~s$^{-1}$ by 5 $R_p$.  The velocity structure is not detectable because the optical depth of the high velocity material is not sufficient, the LOS velocity at the limb of the planet is slower than the radial velocities in general, and because spectral line broadening within the COS instrument smooths the structure out of the line profiles.  If the presence of velocity structure is confirmed in the data \citep{linsky10}, it probably implies that there is detached, optically thick plasma moving at large velocities around the planet.  If this turns out to be the case, interaction with the stellar wind probably plays a role in giving rise to the observed absorption.  Such interaction may also produce turbulence that can broaden the absorption further \citep[e.g.,][]{tian05}.  However, we note that non-thermal broadening such as that proposed by \citet{benjaffel10} does not appear to be necessary to explain the current observations.                                   

\subsection{Ionospheric escape}
\label{subsc:ionescape} 

The escape of heavy atoms and ions has interesting consequences for the nature of the upper atmosphere.  Here we discuss these consequences based on simple analytic arguments, and without explicit use of any complex models. The detection of heavy neutral species can be used to constrain the mass loss rate while the detection of heavy ions outside the atmosphere of the planet potentially constrains the strength of the planetary magnetic field.  For instance, \citet{hunten87} derived an expression for the crossover mass limit $m_c$ for a neutral species $s$ to be dragged along by an escaping neutral species $t$ of mass $m_t < m_s$:
\begin{equation}
m_c = m_t + \frac{k T F_t}{n D_{st} x_t g_0 r_0^2} 
\label{eqn:hunten}
\end{equation}
where $F_t$ is the flux (s$^{-1}$ sr$^{-1}$) of species $t$, $x_t$ is the volume mixing ratio, $g_0$ is gravity at the lower boundary of the model region, and the mutual diffusion coefficient can be roughly estimated from:
\begin{equation}
n D_{st} = 1.52 \times 10^{18} \left( \frac{1}{M_{s}} + \frac{1}{M_t} \right)^{1/2} \sqrt{T} \ \ \ \  \textrm{cm}^{-1} \textrm{s}^{-1} 
\end{equation}
where the masses $M$ are in amu.  We used equation~(\ref{eqn:hunten}) to estimate the mass loss rate that is required to drag neutral O to the exosphere of HD209458b.  Assuming that $x_t \sim$~1, $g_0 =$~10 m s$^{-1}$, $r_0 =$~R$_p$, $T =$~7,200 K, and $n D_{st} =$ 1.3~$\times$~10$^{22}$ m$^{-1}$s$^{-1}$, we obtain $F_t \approx$~2.8~$\times$~10$^{32}$ s$^{-1}$~sr$^{-1}$.  This implies a minimum mass loss rate of 6~$\times$~10$^{6}$ kg~s$^{-1}$.

\begin{figure}
  \centering
  \includegraphics[width=0.7\textwidth]{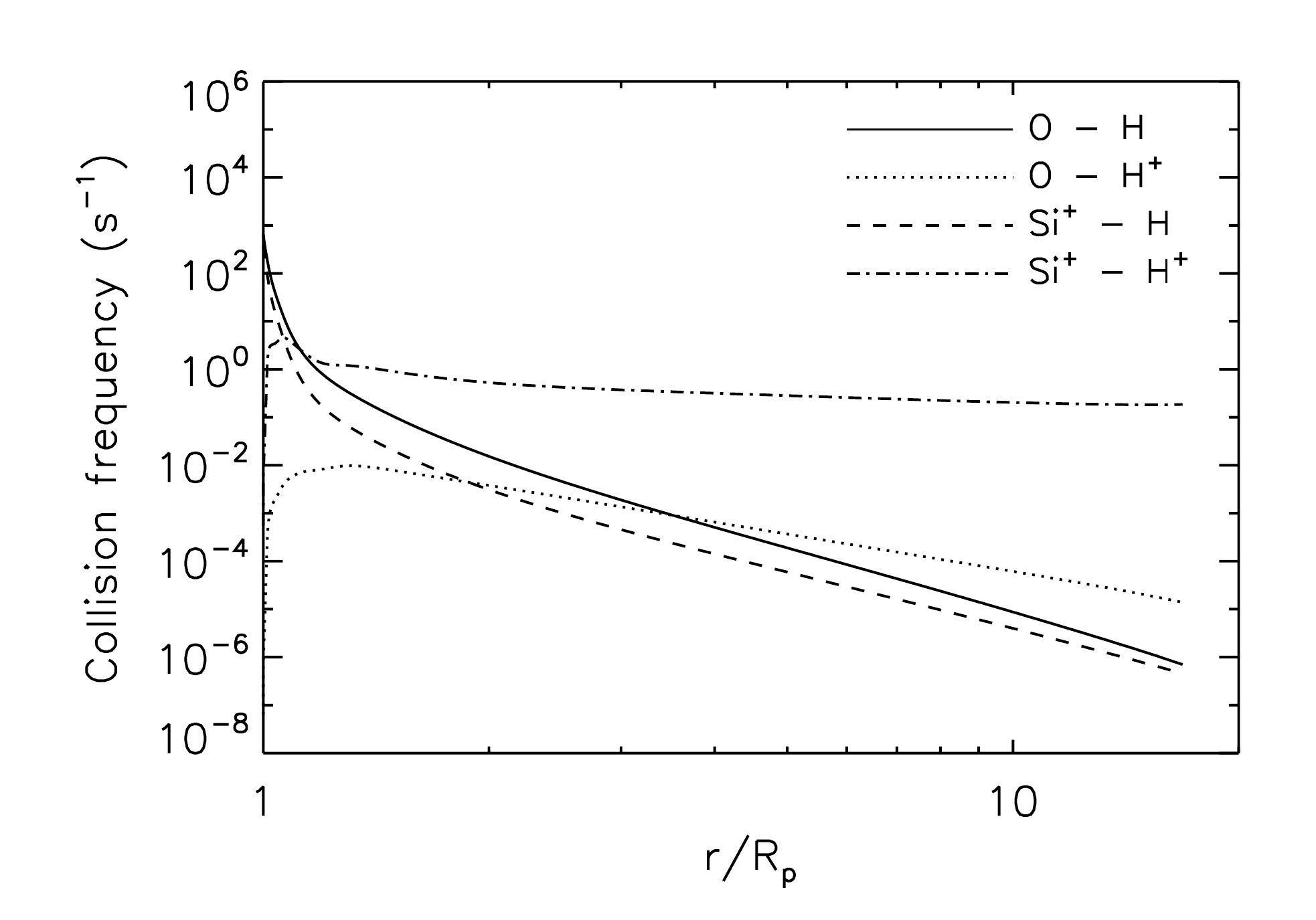}
  \caption{Momentum transfer collision frequencies based on the C2 model.}
  \label{fig:collisions}
\end{figure}

The ionosphere of HD209458b is mostly neutral below 3 $R_p$ but even weak ionization can lead to frequent Coulomb or ion-neutral collisions that enable heavy ions or atoms to escape more efficiently.  In order to illustrate the role of different collisions in transporting O and Si$^+$, Figure~\ref{fig:collisions} shows the collision frequencies for these species with H and H$^+$ as a function of altitude based on the C2 model (Paper I).  We used approximate expressions for resonant and non-resonant ion-neutral collisions, and Coulomb collisions from \citet{schunk00} to calculate the momentum transfer collision frequencies.  The collision frequency between two neutral species, on the other hand, was estimated from the mutual diffusion coefficient as:
\begin{equation}
\nu_{st} = 5.47 \times 10^{-11} \left( \frac{1}{M_s} + \frac{1}{M_t} \right)^{-1/2} \frac{n_t \sqrt{T}}{M_s} 
\end{equation}
where the number density $n_t$ is in $cm^{-3}$.  The results indicate that the transport of O depends on collisions with H below 3.5 $R_p$ whereas the transport of Si$^+$ depends on collisions with H$^+$ at all altitudes.  

Oxygen is the heaviest neutral species detected in the escaping atmosphere, and this implies that the mass loss rate from HD209458b is $\dot{M} >$~6~$\times$ 10$^{6}$ kg~s$^{-1}$.  This result agrees with \citet{vidalmadjar03} although it is less model-dependent and based on different criteria.  The dominance of Coulomb collisions means that the heavy ions can escape even if the mass loss rate is lower than this.  Our models predict mass loss rates of $\dot{M} \approx$~5~$\times$~10$^{7}$ kg~s$^{-1}$ (Table~\ref{table:models}) and thus diffusive separation does not take place in the thermosphere of HD209458b for neutral species with masses less than $\sim$130 amu.  We note that this is the case even if escape is subsonic.  In fact, equation~(\ref{eqn:hunten}) was originally derived for subsonic escape under the diffusion approximation although it is also valid for supersonic escape \citep{zahnle86}.      

We used the collision frequencies to derive expressions for the ion fractions $f_i = n_{\textrm{H}^+}/n_{\textrm{H}}$ at which ion-neutral and Coulomb collisions become important.  The ratio of the non-resonant neutral-ion to neutral-neutral collisions exceeds 10 \% when
\begin{equation}
f_i \approx 10^{-24} \sqrt{ \frac{T}{M_{si} \gamma_s e^2} } 
\end{equation}
where $i$ denotes H$^+$, $s$ denotes the colliding species, $\gamma_s$ is the neutral polarizability and all units are in cgs.  However, the collision of O with H$^+$ is resonant and in this case the required ratio differs slightly from the above expression.  The ratio of the Coulomb to non-resonant ion-neutral collisions, on the other hand, exceeds 10 \% when
\begin{equation}
f_i \approx 4.24 \times 10^{11} \frac{T^{3/2}}{Z_s^2 Z_i^2} \sqrt{ \frac{M_{sn} \gamma_n e^2}{M_{si}} }. 
\end{equation}
where $i$ denotes $H^+$ (or the dominant ion) and $n$ is $H$ (or the dominant neutral).  For Si$^+$ this fraction is $f_i \approx$~10$^{-4}$ (with $\gamma_{\textrm{H}} =$~6.7~$\times$~10$^{-25}$ cm$^3$).  These equations can be used to determine if equation~(\ref{eqn:hunten}) is valid, or if more complex plasma models are required.   

\citet{trammell11} argued that HD209458b could have a strong planetary magnetic field that can impede the escape of ions from equatorial regions and restrict it to the polar regions.  Although the magnetic field does not directly interfere with the escape of the neutral atoms, the trapped ions can stop them from escaping if the neutral-ion collision frequency is sufficiently high.  Unfortunately, transit observations are not spatially resolved and they cannot be used directly to determine if escape is limited to the poles, or if the atmosphere is also escaping over the equator.  However, the transit depths depend on the size of the optically thick obstacle covering the star.  If mass loss is suppressed at low and mid-latitudes, the heavy species are no longer mixed into the upper atmosphere other than at the poles where they are allowed to escape.  This means that the cross-sectional area covered by the ions shrinks and may become insufficient to explain the observations.  Even if the plasma spreads to cover a larger area after being ejected from the poles, it is diluted in the process and thus it is not clear if the resulting cloud would have sufficient optical depth to be detectable.  

Assuming that the charged particles escape the Roche lobe of the planet unimpeded at the equator, we estimated an upper limit for the planetary magnetic field by evaluating the magnetic moment that allows them to do so.  In order to estimate this limit, we calculated the plasma $\beta$ and the inverse of the second Cowling number (Co$^{-1}$) below 5 $R_p$ in the C2 model from:
\[ \beta = \frac{2 \mu_0 p}{B^2} \ \ , \ \ \textrm{Co}^{-1} = \frac{\mu_0 \rho v^2}{B^2} \]
where $v$ is the vertical velocity.  Assuming a dipolar magnetic field, we obtained a limiting magnetic moment of 3.2~$\times$~10$^{25}$ Am$^2$ or 0.04 $m_J$ ($m_J =$ 1.5~$\times$10$^{27}$ Am$^2$ is the magnetic moment of Jupiter).  This magnetic moment ensures that $\beta >$~10 below 5 R$_p$ and that Co$^{-1}$ reaches 10 by 5 $R_p$.  Magnetic moments of $m_p \lesssim 0.04~m_J$ agree quite well with the scaling laws discussed by \citet{griesmeier04}.  We note that the limiting moment produces an equatorial surface field that is approximately 4 times lower than the surface field of the Earth.       

\subsection{Cloud formation on HD209458b}
\label{subsc:clouds}

We have shown that a substantial abundance of silicon in the upper atmosphere is required to produce a detectable transit in the Si III line.  If the silicon ions originate from the atmosphere of the planet, at least a solar Si/H ratio is necessary.  According to thermochemical equilibrium models, silicon should condense into clouds of forsterite (Mg$_2$SiO$_4$) and enstatite (MgSiO$_3$) in the lower atmosphere of HD209458b \citep{visscher10}.  If the formation of enstatite is suppressed, silicon should condense to form quartz (SiO$_2$) instead.  In any case, condensation is expected to remove almost all of the silicon from the upper atmosphere.  The detection of Si$^{2+}$ implies that the abundance of silicon in the thermosphere is substantial, and thus the condensation of silicon does not take place in the atmosphere of HD209458b.  This has significant implications for the structure and dynamics of the atmosphere.       

\citet{sing08a,sing08b} analyzed the absorption line profile of Na in the atmosphere of HD209458b in detail and argued that the abundance of Na is depleted above the 1 mbar level.  They suggested that this is due to the condensation of sodium into Na$_2$S, although ionization could not be ruled out decisively.  Based on the condensation temperature of Na, they argued that the temperature in the upper atmosphere of HD209458b near 1 mbar is 420~$\pm$~190 K.  This temperature is significantly lower than the outcome of typical radiative transfer models for HD209458b \citep[e.g.,][]{showman09}.  We note that the condensation temperature of forsterite and enstatite is higher than 1,300 K at 1 mbar \citep{visscher10}.  Because silicon clouds do not form, the Na$_2$S clouds cannot form either.  Further, the formation of Na$_2$S relies on H$_2$S, which is dissociated above the 1 mbar level \citep{zahnle09}.  This implies that any depletion of Na at high altitudes is most likely due to photoionization and/or thermal ionization.    

\begin{figure}
  \centering
  \includegraphics[width=0.7\textwidth]{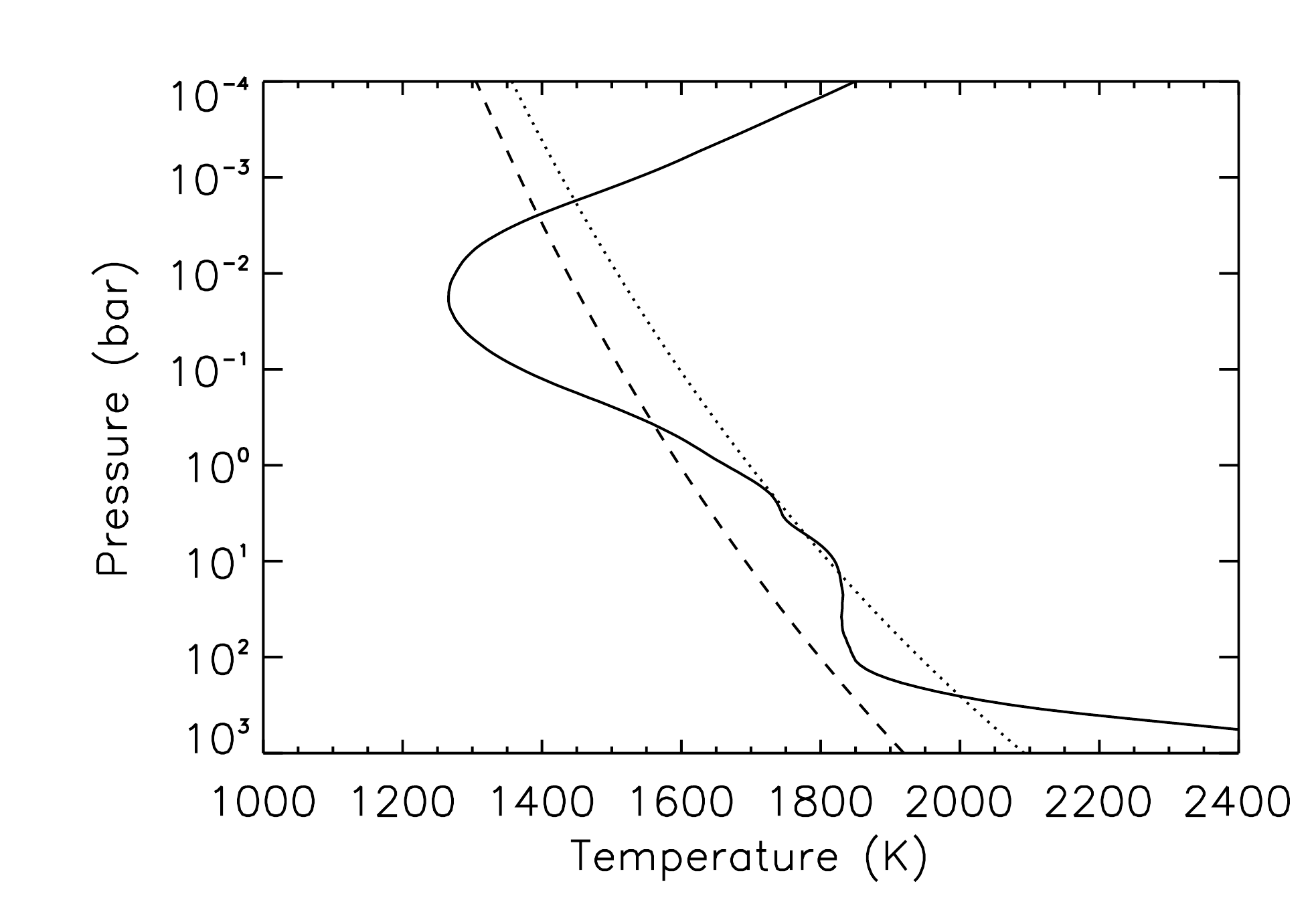}
  \caption{The dayside temperature-pressure (T-P) profile of HD209458b from \citet{showman09} that includes a temperature inversion at low pressures.  The condensation curves for forsterite (dotted line) and enstatite (dashed line) are shown.}
  \label{fig:condensation}
\end{figure}

Figure~\ref{fig:condensation} shows a dayside temperature (T-P) profile for HD209458b based on \citet{showman09}.  This profile is similar to the dayside temperature profile adopted by \citet{moses11}.  The figure also shows the condensation curves for forsterite and enstatite.  The T-P profile crosses the condensation curve for forsterite below the 100 bar level.  However, the temperature profile in the deep atmosphere is uncertain, and the current profile only barely crosses the condensation curve.  Also, the formation of forsterite ties only a fraction of the total abundance of silicon into clouds \citep{visscher10}.  However, the T-P profile crosses the condensation curves for both forsterite and enstatite in a `cold trap' near 10 mbar.  To prevent this, the temperature in the cold trap would have to be $T \gtrsim$~1,600 K.  This is not totally unbelievable but probably unlikely.  In any case, the temperature is close enough to the condensation curves so that moderate vertical transport might be able to preserve silicon above the cold trap.

\citet{spiegel09} explored a range of turbulent diffusion coefficients $K_{zz}$ that would be required to prevent the condensation of TiO and VO in the atmospheres of different EGPs, including HD209458b.  In fact, they assumed that condensates form in the cold trap but are then transported to higher altitudes where they evaporate.  Their work ignores the detailed chemistry of condensation, and thus the results are simply based on the ratio of $K_{zz}$ to the diffusion coefficient estimated from
\[ D_p \approx v_p H \]
where $v_p$ is the particle settling velocity and $H$ is the pressure scale height.  The formation of condensates is probably too complicated for such a simplistic treatment, but the results provide some guidance on the value of $K_{zz}$ that is required to lift the condensates from the cold trap.

We calculated $D_p$ for forsterite grains with a radius of 0.1 $\mu$m and density\footnote{This is the density of the material in the particles, not the density of the particles in the atmosphere.} of 3,200 kg~m$^{-3}$ \citep{fortney03,cooper03}.  The settling velocity for such particles in the cold trap is $v_p \approx$~3~$\times$~10$^{-3}$ m~s$^{-1}$ and $D_p \approx$~2~$\times$~10$^3$ m$^2$~s$^{-1}$.  Given that the upper edge of the cold trap is near the 1 mbar level where $D_p$ is higher, $K_{zz} \gtrsim$~10$^5$ m$^2$~s$^{-1}$ is sufficient to prevent the settling of the cloud particles in the lower atmosphere.  We note that mass loss does not help to enhance the mixing of the particles at low altitudes significantly.  The vertical velocity based on the mass loss rate of 10$^7$ kg~s$^{-1}$ is only 4.8~$\times$~10$^{-7}$ m~s$^{-1}$ at 10 mbar and 8~$\times$~10$^{-3}$ m~s$^{-1}$ at 1 $\mu$bar.                  

Estimating $K_{zz}$ on extrasolar planets is very difficult, partly because there is no agreement on exactly what physical processes this parameter describes even in much more sophisticated solar system applications.  The most recent estimates for HD209458b are based on assuming that $K_{zz} \sim \overline{v} H$ where $\overline{v}$ is the rms vertical velocity from circulation models \citep[e.g.,][]{showman09}.  Based on the GCMs of \citet{showman09} and an assumed density dependence with altitude, \citet{moses11} estimated that the high pressure value of $K_{zz} =$~10$^6$ m$^2$~s$^{-1}$, which implies that $K_{zz} \gtrsim$~10$^7$ m$^2$~s$^{-1}$ at $p \lesssim$~10 mbar.  If such high values are realistic, turbulent mixing is almost certain to prevent the settling of the condensates and to preserve gaseous silicon in the upper atmosphere.     

We note that the above requirements on the value of $K_{zz}$ may in fact be overestimated because they are based on the assumption that clouds particles form in the cold trap.  Cloud formation has to be studied in the context of a photochemical model that includes the chemistry of condensation.  If the chemical timescale is longer than the transport timescale, the cloud droplets may not form in the first place before the gas escapes from the cold trap.  Also, the temperature structure near the cold trap needs to be constrained in greater detail.  The formation of condensates is a complex problem that will be studied in future work (Lavvas et al., \textit{in preparation}) in order to better constrain the required values of $K_{zz}$.  For our purposes it is sufficient to note that the current estimates of $K_{zz}$ imply, in agreement with the observations, that silicon clouds do not form in the upper atmosphere of HD209458b.   

\section{Discussion and Conclusions}
\label{sc:discussion} 

We have used multiple observational constraints and theoretical models to characterize the upper atmosphere of HD209458b.  Contrary to many of the earlier studies, we did not treat the thermosphere independently of the rest of the atmosphere.  In fact, we have shown that observations of the upper atmosphere can be used to obtain useful constraints on the characteristics of the lower atmosphere.  This is important because the extended upper atmospheres of close-in EGPs produce much larger transit depths than those arising from the lower atmosphere.  In this work, we concentrated mostly on the FUV transit measurements.  Future work should explore the possibility of extending the range of possible observations to other wavelength regions, as well as obtaining repeated observations in the FUV lines.  Theoretical models should be developed to support new observations and to clarify the interpretation of the existing data.         

In agreement with K10, we showed that the H Lyman $\alpha$ transit observations \citep{vidalmadjar03,vidalmadjar04,benjaffel07,benjaffel08,benjaffel10} can be fitted with a layer of H in the thermosphere that is described by three simple parameters.  These are the pressure at the bottom of the H layer, the mean temperature in the thermosphere, and a cutoff level due to ionization.  The most important parameters are the pressure at the lower boundary and the mean temperature.  Because H is the dominant species in the thermosphere, the data can be used to estimate the temperature of the thermosphere.  Choosing a lower boundary pressure of 1 $\mu$bar based on the location of the H$_2$/H dissociation front in recent photochemical models \citep{moses11} and observational constraints \citep{france10}, we measured a mean temperature of about 8,250 K in the thermosphere below 3 $R_p$.  However, the uncertainty of the observations is large, and the 1$\sigma$ upper and lower limits on this temperature are approximately 6,000 K and 11,000 K, respectively\footnote{This uncertainty does not include the possible uncertainties in the other parameters of the fit.}.  

We used a hydrodynamic model that treats the heating of the upper atmosphere self-consistently and the average solar XUV spectrum (Paper I) to show that a mean temperature of 8,250 K in the upper atmosphere below 3 $R_p$ is higher than the maximum temperature allowed by stellar heating.  Given that a net heating efficiency equal to or higher than 100 \% is unrealistic, this temperature requires either a non-radiative heat source, additional opacity, or it implies that the XUV flux of HD209458 is higher than the corresponding solar flux.  Interestingly, this would also imply that the mass loss rate could be higher by a factor of 2 or more than previously anticipated (e.g., Paper I).  However, the uncertainty in the H Lyman $\alpha$ observations also allows for a slightly lower temperature of 7,200 K that is typical of stellar heating based on the average solar XUV flux and our best estimate of the net heating efficiency.  Therefore the temperature implied by the H Lyman~$\alpha$ observations and the mean temperature of the basic hydrodynamic models are in good agreement.  

We note that the temperature in the lower thermosphere near the 1 $\mu$bar region has been constrained previously by \citet{vidalmadjar11a,vidalmadjar11b} who used the Na D lines to constrain the density and temperature profiles in the atmosphere of HD209458b.  Their results point to a temperature of $\sim$3,600 K that is actually higher than the temperature at the lower boundary of our hydrodynamic model and consistent with a high mean temperature at lower pressures in the thermosphere.  However, we caution the reader that the temperature profile based on the Na D lines may not be accurate.  This is because \citet{vidalmadjar11a} used a simple expression for the optical depth due to Na that is based on the scale height of the atmosphere [their equation (1)].  This expression is only valid if Na is uniformly mixed with H$_2$.  Since the authors also argue that Na is depleted (i.e., its mixing ratio changes with altitude) above $\sim$10 mbar, its density scale height cannot be used to estimate temperatures accurately.   

The detection of O in the thermosphere allowed us to constrain the mass loss rate from HD209458b based on the crossover mass concept formulated by \citet{hunten87}.  This is because O is transported to high altitudes in the upper atmosphere primarily by collisions with H.  We found that a minimum mass loss rate of 6~$\times$~10$^6$ kg~s$^{-1}$ is required to prevent the diffusive separation of O.  This is one of the most reliable constraints on the mass loss rate available at present.  Our hydrodynamic calculation based on the average solar XUV flux predicts a mass loss rate close to 5~$\times$~10$^7$ kg~s$^{-1}$.  This implies that species with a mass up to $\sim$130 amu are uniformly mixed in the thermosphere.  Similar constraints do not apply to heavy ions.  They are transported to high altitudes by Coulomb collisions with H$^+$ that are much more efficient in preventing diffusive separation compared to collisions of neutral atoms with H. 

In agreement with K10, we found that the presence of O with a solar abundance and a temperature based on the H Lyman $\alpha$ measurements explains the transits observed in the O I lines.  Our models predict a line-integrated transit depth of approximately 4 \% that deviates from the measured transit depth by 1.5$\sigma$ and thus agrees with the uncertainty of the observations.  However, the predicted transit depths fall systematically short of the measured value.  K10 suggested that a higher transit depth is possible if the O/H ratio is supersolar, the temperature of the thermosphere is higher than expected, and/or the observations probe escaping gas outside the Roche lobe of the planet, and we agree with their conclusions.  As we explain below, we found that the same is true for the other heavy species.    

In order to calculate predicted transits in the C II and Si III lines, we created model emission line profiles for HD209458 based on SUMER observations of the Sun \citep{curdt01} adjusted to the observations of HD209458 \citep{linsky10}.  We did not find evidence for significant absorption by the ISM in the C II 1335.7~\AA~or the Si III line.  Parts of the C II 1334.5~\AA~line, on the other hand, are optically thick in the ISM and we took this into account in our models.  We note that resolved observations of the emission line profiles can be used to characterize the composition of the ISM and the activity of the host star.  This information is a valuable byproduct of the FUV transit observations that typically need to be repeated several times.

With solar abundances, the same models that agree with the H Lyman~$\alpha$ transit observations tend to underestimate the transit depths in the C II and Si III lines.  Similarly with the O I lines, higher transit depths in the C II and Si III lines are possible if the mean temperature of the absorbers is higher than expected and/or the C/H and Si/H ratios are supersolar.  With solar abundances, a 1$\sigma$ agreement with the observed line-integrated transit depths is possible if the stellar XUV flux (or the stellar flux combined with an additional heat source) is higher than or equal to 2 times the average solar XUV flux.  This corresponds to typical solar maximum conditions, and it is quite interesting that a similar enhancement may be required to explain the relatively high mean temperature in the thermosphere that we estimated earlier.  Alternatively, with heating based on the average solar flux, an agreement with the observations is possible with the O/H, C/H and Si/H enhanced by a factor of $\sim$3--5 relative to solar abundances \citep{lodders03}.  In any case, the atmosphere is escaping with a minimum mass loss rate given above.  This is evident because of both the high temperature of the thermosphere and the detection of heavy species at high altitudes.       

We note that the transit observations are affected by stellar variability that can lead to significant changes in the observed transit depths.  Spatial variations of intensity on the stellar disk during maximum activity or limb brightening may render the transit occasionally undetectable, or enhance it by a significant factor.  Generally, observations obtained during periods of minimum activity are more reliable.  \citet{benjaffel07} characterized the short-term variability of HD209458 in the H Lyman $\alpha$ line, but the variability of the O I, C II, and Si III are poorly characterized.  This introduces an additional element of uncertainty into the transit depths that needs to be constrained by repeated observations of the star and the transits in the FUV lines.  It also means that the qualitative explanation of the present observations that is based on heating by the average solar XUV flux and solar abundances (Paper I) may be sufficient even if the predicted transit depths in the O I, C II, and Si III lines do not exactly match the current measurements.  On the other hand, if higher transit depths are confirmed, they can be used to further constrain the mean temperature and abundances as we have shown.    

The detection of heavy ions escaping the atmosphere can potentially be used to constrain the magnetic field strength of the planet.  We estimated an upper limit of 0.04 $m_J$ for the magnetic moment that allows the heavy ions to escape unimpeded at the equator.  The estimated magnetic moment agrees with the scaling laws for the magnetic field strengths of tidally locked close-in EGPs by \citet{griesmeier04}.  On the other hand, a strong magnetic field inhibits the flow of ions from the equator and may only allow for escape at the poles \citep[e.g.,][]{trammell11}.  If the neutral-ion collision frequencies are sufficient, the trapped ions may also prevent the neutral atoms from escaping.  We note that a uniform upward flux is required to mix the atmosphere, and escape at the poles may not be sufficient to create a large enough obstacle to explain the transits in the O I, C II, and Si III lines.  Detailed models of the magnetosphere that include the heavy species are required to assess if this is the case or not.     

The detection of Si$^{2+}$ in the upper atmosphere means that silicon cannot condense to form enstatite, forsterite, or other condensates in the lower atmosphere.  This is clear because at least a solar abundance of silicon in the thermosphere is required to explain the large transit depth in the Si III line.  According to the calculated temperature profiles for HD209458b \citep[e.g.,][]{showman09}, condensation is expected in a cold trap near the 10 mbar level.  Provided that the temperature is not much higher than expected near the cold trap, efficient mixing is required to prevent condensation.  Following an argument similar to that of \citet{spiegel09}, we estimated that an eddy mixing coefficient of $K_{zz} \gtrsim$~10$^5$~m$^2$~s$^{-1}$ below 1 mbar is sufficient to prevent the condensation of forsterite and enstatite in the cold trap.  We note that much higher values than 10$^5$ m$^2$~s$^{-1}$ were assumed by recent photochemical models by \citet{garciamunoz07} and \citet{moses11}.  

A stratospheric temperature inversion may also be necessary to suppress condensation.  This is because the cold trap cannot extend to much lower pressure than 1 mbar or the required values of $K_{zz}$ become too high.  Also, the lack of condensation implies that the temperature of the lower atmosphere should be relatively high.  The detection of silicon in the upper atmosphere thus provides further evidence for a stratosphere on HD209458b that was first proposed by \citet{knutson08}.  We note that existing radiative transfer models do not account for molecular opacity at UV wavelengths or visible absorbers potentially generated by photochemistry \citep[e.g.,][]{zahnle09}.  Our results provide motivation for more detailed models of thermal structure below the thermosphere that can constrain the chemistry of the lower atmosphere.  Once this is achieved, better constraints on the dynamics can be derived.

We also address an old problem related to the atmosphere of HD209458b.  Based on the observed transits in the Na D lines, \citet{charbonneau02} argued that Na is depleted in the atmosphere because their cloudless solar composition model predicted significantly deeper absorption in the D lines.  In addition to photoionization, molecular chemistry, and low primordial abundance of Na, they suggested that the formation of high altitude clouds can explain the observed depletion.  Later \citet{sing08a,sing08b} found further evidence for the depletion of Na above the 3 mbar level, and argued that condensation of Na$_2$S is the most likely explanation.  The fact that silicon does not condense implies that condensation of Na$_2$S is also unlikely.  The observed depletion is therefore most likely due to photoionization and/or thermal ionization.  However, the density profile and ionization state of Na should be studied in the context of molecular and ion chemistry below the 0.1 $\mu$bar level to verify that this is the case.\\
\\  

We thank H. Menager, M. Barthelemy, N. Lewis, and D. S. Snowden for useful discussions and correspondence, and A. Showman and N. Lewis for sharing some of their temperature profiles.  We also acknowledge the "Modeling atmospheric escape" workshop at the University of Virginia and the International Space Science Institute (ISSI) workshop organized by the team "Characterizing stellar and exoplanetary environments" for interesting discussions and an opportunity to present our work.  The calculations for this paper relied on the High Performance Astrophysics Simulator (HiPAS) at the University of Arizona, and the University College London Legion High Performance Computing Facility, which is part of the DiRAC Facility jointly funded by STFC and the Large Facilities Capital Fund of BIS.  SOLAR2000 Professional Grade V2.28 irradiances are provided by Space Environment Technologies.

\end{document}